\documentclass[12pt,usenames,dvipsnames,svgnames,table]{article}
\usepackage{graphicx}
\usepackage{bm}
\usepackage{amsmath}
\usepackage{amsfonts}
\usepackage{amssymb}
\usepackage{tikz}
\usepackage{epstopdf}
\usepackage{epsfig}
\usepackage{comment}
\usepackage{setspace}
\usepackage{url}
\usepackage{color}
\usepackage{float}
\usepackage{xcolor}
\usepackage[authoryear,round]{natbib}
\usepackage{rotating}
\usepackage{geometry}
\usepackage{pdflscape}
\usepackage{longtable, booktabs, tabularx}
\usepackage{caption}
\usepackage{subfig}
\usepackage{dsfont}
\usepackage{hyperref}
\hypersetup{
  colorlinks   = true, 
  urlcolor     = blue, 
  linkcolor    = blue, 
  citecolor   = blue, 
}
\newtheorem{proposition}{Proposition}
\newtheorem{corollary}{Corollary}

\newtheorem{example}{Example}

\newcommand{\dd}{{\rm d}}

\newcolumntype{L}[1]{>{\arraybackslash}p{#1}}
\allowdisplaybreaks

\begin{document}
\title{Entry and disclosure in group contests\thanks{We thank three anonymous reviewers for their insightful comments, as well as participants of the 5th annual conference on \emph{Contests: Theory and Evidence} and the 95th annual meeting of the \emph{Southern Economic Association}. All the remaining errors are ours.}}
\author{Luke Boosey\thanks{Department of Economics, Florida State University, Tallahassee, FL 32306-2180, USA; e-mail: \texttt{lboosey@fsu.edu}} \quad Philip Brookins\thanks{Department of Economics, University of South Carolina, Columbia, SC 29208, USA; e-mail: \texttt{philip.brookins@moore.sc.edu}} \quad Dmitry Ryvkin\thanks{Department of Economics,  School of Economics, Finance and Marketing, RMIT University, Melbourne, VIC 3000, Australia; e-mail: \texttt{d.ryvkin@gmail.com}}}
\date{\small This version: \today}

\maketitle

\begin{abstract}
\noindent We study information disclosure policies for contests among groups. Each player decides whether or not to participate in competition as a member of their group. We consider a generalized all-pay auction setting, in which within-group aggregation of effort is best-shot and the group with the highest performance wins the contest. Players' values for winning are private information at the entry stage, but may be disclosed at the competition stage. We compare three disclosure policies: (i) no disclosure, when the number of entrants remains unknown and their values private; (ii) within-group disclosure, when this information is disclosed within each group but not across groups; and (iii) full disclosure, when the information about entrants is disclosed across groups. For contests between individuals, information disclosure always reduces expected aggregate investment. However, this is no longer true in group contests: Within-group disclosure unambiguously raises aggregate investment, while the effect of full disclosure is ambiguous.



\bigskip



\end{abstract}

\newpage

\onehalfspacing
\section{Introduction}

Across a wide range of economic, political, and social environments, competition takes place between groups of individuals who align themselves in pursuit of some common goals. For example, many interest groups engaged in lobbying activities consist of diverse collections of individuals with broadly aligned objectives. Within organizations, managers may solicit project submissions from multiple divisions or teams of employees and reward the team with the best proposal. Crowdsourcing R\&D platforms such as the XPRIZE Foundation invite groups to compete for prizes solving complex innovation problems. Such competitive settings can be broadly categorized as \emph{group contests}. In each group, individuals invest effort or other resources to increase their group's performance, which in turn improves the group's chances of winning a valuable prize.

In this paper, we study competition between groups in which the decision to enter competition as a member of a particular group is endogenous. In our setting, there are a fixed number of competing groups, each with a pool of \emph{potential} participants. In an initial entry stage, each player decides whether or not to participate in competition as a member of their group. Participants must forgo some outside option or, equivalently, face a cost of entry. Then, in a second stage, participants make investment or effort decisions that determine the group's performance level. We analyze and compare three \textit{disclosure policies} that dictate the information available to participants at the time they make their investments.

We consider activities in which group performance is determined solely by the \emph{best-shot}, i.e., the maximum investment chosen by a group member \citep{Baik-Shogren:1998,Chowdhury-et-al:2013}.\footnote{Other popular approaches are to model the group performance level using perfect substitutes aggregation technology \citep{Katz-et-al:1990,Baik:1993,Baik:2008,Baik-et-al:2001,Munster:2009} or weak-link (perfect complements) technology \citep{Lee:2012}. \citet{Kolmar-Rommeswinkel:2013} and \cite{Brookins-et-al:2015} consider varying degrees of complementarity with a CES technology.} Such a setting may arise, for example, in lobbying environments where the official who is lobbied considers only the individual agents on each side of an issue who made the most compelling case; or within an organization where a team of employees pitches only the most promising project idea to the management. One can think also of a market setting where a client, such as a real-estate developer, solicits projects from multiple firms, and each firm conducts an internal selection process and presents its best proposal. 

We model group competition as a generalized all-pay auction 
in which the group with the highest performance (the highest best-shot) wins the contest with certainty.\footnote{This setting corresponds to a perfectly discriminating contest success function (CSF). Alternative environments, in which the contest is imperfectly discriminating, include those with a lottery CSF of \cite{Tullock:1980} and its generalizations. In these settings, the group with the highest performance has a higher probability of winning, but does not win with certainty.} Moreover, we assume that the prize associated with winning is a group-specific public good. Thus, the value of winning for each member of the winning group is equal to her individual private value, regardless of the size of the group.\footnote{Note that \emph{potential members} of the winning group who do not participate do not receive any benefit from the prize being awarded to the group they could have joined. This is the case, for example, for researchers deciding whether or not to join their colleagues on a grant application, politicians joining various caucuses or factions, or (potential) plaintiffs in group litigation.} Together, these features of the environment generate incentives for individuals within groups to free-ride on the investments made by their fellow group members. These free-riding incentives underscore one of the key differences between contests among groups and contests among individuals. As such, the environment we consider provides a rich and previously unexplored interplay between endogenous entry, free riding and information disclosure in contests.

The game proceeds in two stages. At the initial entry stage, players' values for the prize are private information, although it is common knowledge that they are drawn independently from the same commonly known distribution. Depending on the disclosure policy, information about the number of participating group members and their values may become known to participants at the beginning of the second stage, prior to their investment decisions.\footnote{We assume this disclosure of information is exogenous. It may be the effect of explicit design decisions, for example, by a contest sponsor; or it may be a result of naturally occurring dissemination of information due to environmental factors, such as the spatial or network structure of agents' interactions.} In the second stage, participants simultaneously choose an irreversible, costly investment, and the outcome of the contest is determined. 

We first consider, as a benchmark, the case of contests among individuals and show that expected aggregate investment is always lower when information about others' types is disclosed to those who enter. In this case, equilibrium entry is independent of disclosure. However, equilibrium aggregate investment corresponds to the (truncated) expectation of the second highest valuation without disclosure; and to the sum of the expected bids by the two highest valuation entrants in a complete information all-pay auction, with disclosure. The latter entails an efficiency loss because the lower valuation player can win, and a loss of revenue because the lower valuation player can bid zero, with a positive probability. There is also no scope for free riding or a coordination problem disclosure may alleviate; thus, nondisclosure leads to a larger investment.

We then turn to contests among groups and consider three different disclosure rules. In the \textbf{no disclosure} (ND) setting, all entry decisions and valuations remain private information throughout the investment stage. As a result, entrants face incomplete information about the number of entrants both within their own group and in other groups, as well as about others' valuations for the prize. While this setting is stylized and serves primarily as a benchmark, it approximates contexts such as organizations with minimal interaction among personnel (e.g., telecommuting workers or employees on disjoint schedules). Another example is lobbying---either in politics, where lobbyists on opposing sides may be unaware of the number and types of other aligned or opposing lobbyists; or within organizations, where multiple employees privately lobby a manager to influence a particular decision.\footnote{Note that the ND setting also does not allow for communication such as cheap talk, which makes coordination within groups difficult.} In the \textbf{full disclosure} (FD) setting, entrants are informed, prior to making investment decisions, about the number and valuations of all entrants. This condition is applicable, for example, to online crowdsourcing competitions (e.g., TopCoder.com and Kaggle.com) whereby ``leaderboards'' continuously and publicly display team information, such as the number of team members and their skill levels, player bios and past accomplishments.\footnote{See, e.g., the following leaderboard on Kaggle.com: \url{https://www.kaggle.com/c/data-science-bowl-2018/leaderboard}.}   In the third setting, which we call \textbf{within-group disclosure} (WD), entrants are informed about the number and valuations of all entrants within their own group, but do not learn any information about the entrants in other groups \citep{Brookins-Ryvkin:2016}.\footnote{Of course, there are many possible ways to implement partial disclosure. For example, one can consider a setting where the number of entrants in one's own group is known, but their types are unknown; or a setting where the numbers of entrants in other groups are known but their types are unknown. While potentially interesting, such models break symmetry at the bidding stage to the extent that makes them intractable.} This information structure is found in many naturally occurring competitive settings. For example, when the US Department of Defense solicits prototype designs for new equipment from multiple defense contractors (such as Boeing or Lockheed Martin), competitor information is strictly walled off between bidders. However, the internal competition between different research or engineering units at a particular contractor is generally transparent, facilitating within-group coordination. There are other ``blind'' competitive settings that only announce competitor information (e.g., background, performance, experience and ability) at the end of the contest stage, such as competition for federal research grants. The importance of the WD information structure is most apparent in such settings where the designer restricts information between competing groups to maintain competitive tension, while encouraging transparency \textit{within} groups to facilitate coordination on the most promising ``best-shot."

While the ND and FD settings are standard in the literature, within-group disclosure is less frequently considered, despite being highly relevant for contests between groups. In fact, in group contests, the case where participants are well-informed about their fellow group members and uninformed about others is arguably a more natural benchmark than the ND setting.

The three disclosure rules we consider allow us to study the complex interplay between three different effects in group contests with endogenous entry: \emph{efficiency}, \emph{free riding}, and \emph{entry}. The ND rule is efficient (in that the highest valuation bidder wins with probability one), but, due to lack of coordination, there is free riding within groups. The WD rule remains efficient and solves the coordination problem. The FD rule solves the coordination problem but it is inefficient; however, FD also induces more entry than both ND and WD.

Formalizing these insights, we show that expected aggregate investment is unambiguously \emph{higher} under the WD rule as compared to the ND rule. Equilibrium entry does not differ between the two settings, and the reduction in free riding is sufficiently strong so that the expected aggregate investment supplied by the highest types in each group in WD exceeds not only the same measure (expected aggregate investment) but even the expected sum of \textit{all} entrants' investments in the ND case.

Turning to full disclosure (FD), we show that the effect of disclosure on expected aggregate investment in contests among individuals is often \emph{reversed} in contests between groups. While in individual contests FD unambiguously leads to a lower expected aggregate investment than ND (or WD, which is equivalent to ND in that case), in group contests aggregate investment may be higher or lower in the FD setting than under either of the other two disclosure rules. When entrants are informed about the number and types of entrants in all groups, the contest collapses to an individual all-pay auction of complete information among the groups' ``leaders.'' In this setting, we first prove that there is more entry under FD than under ND and WD. However, as in a typical all-pay auction equilibrium, at most two highest types among the group leaders actively invest with positive probability. As a consequence, it is possible for expected aggregate investment to be lower under FD than under ND. Nevertheless, provided the distribution of types is sufficiently elastic and the (expected) group size is large enough, the distributions of the top two order statistics among the groups' best entrants are shifted towards the upper bound of the type space, which leads the two active entrants to invest more aggressively. As a result, when the distribution of types satisfies this (sufficient) elasticity condition, full disclosure also increases expected aggregate investment compared to the setting with no disclosure. 

The rest of the paper proceeds as follows. In Section \ref{sec:lit}, we review related literature. Section \ref{sec:model} describes the model. The benchmark case of contests among individuals is considered in Section \ref{sec:individuals}, and our main results and numerical illustrations for group contests are presented in Section \ref{sec:groups}. Section \ref{sec:extensions} presents several extensions, and Section \ref{sec:conc} concludes. All missing proofs are collected in Appendix \ref{sec:appendix}.

\section{Related literature}\label{sec:lit}

A fundamental difference between group contests and individual contests is that players in the former typically face some incentive to free-ride on the investments made by other group members. In general, the impacts of such free-riding incentives depend on the size of the group, the type of within-group investment aggregation technology, and on the information available to the participants. Moreover, informational conditions may affect the decision to participate in the first place. Recent years have seen a revival of interest in contests with endogenous entry and in the effect of disclosing the number of entrants. However, virtually all of the existing research focuses on contests among individuals. As such, one of our goals in this paper is to explore the interaction between endogenous entry and free-riding incentives that is unique to group contests.

Our general setting is closely related to the literature on contests in which the prize is a group-specific public good \citep[see, e.g.,][]{Katz-et-al:1990,Baik:1993,Chowdhury-et-al:2013, Kolmar-Rommeswinkel:2013,Barbieri-Malueg:2016, Eliaz-Wu:2018,Barbieri-et-al:2019}.\footnote{There is also a large literature on group contests where the prize is a private good that must be divided between the members of the winning group. For example, see \citet{Nitzan:1991,Lee:1995, Skaperdas:1998,Warneryd:1998,Konrad-Leininger:2007,Munster:2007} and \citet{Nitzan-Ueda:2009,Nitzan-Ueda:2011}.} Only a handful of papers in this literature consider the setup with private information \citep{Fu-et-al:2015,Barbieri-Malueg:2016,Brookins-Ryvkin:2016,Eliaz-Wu:2018,Barbieri-et-al:2019}. 

The baseline features of our group contest environment are most similar to \citet{Barbieri-Malueg:2016}. As in their model, we consider a setting where players' values for winning are private information, group performance is determined by the ``best-shot'' of its members, and the highest performing group wins the contest. The focus of the analysis in \citet{Barbieri-Malueg:2016} is on the comparative statics of individual, group, and aggregate investment, and of the equilibrium probability of winning, as the size of the groups and the number of groups are varied. In contrast, our goals in the current paper are to understand competition when the decision to join a group is endogenous, and to compare equilibrium behavior across different information disclosure policies. To this end, we incorporate into the model two new features. First, we allow for endogenous entry by potential group members prior to the investment stage, which has previously been studied only in contests and auctions among individuals. Second, we vary the information disclosure that takes place between the entry stage and the investment stage regarding group size and participants' types. The impact of information disclosure within groups has been analyzed by \cite{Barbieri-et-al:2019} in group contests with weakest-link aggregation. They show that via cheap-talk communication within groups, higher levels of effort can be achieved due to improved coordination. This result is similar to how within-group disclosure helps solve the coordination problem in our WD setting.

While our study is, to the best of our knowledge, the first to consider endogenous entry into competing \textit{groups}, there is substantial previous work on \textit{individual} contests and auctions with endogenous entry. In all-pay contest environments, key theoretical insights regarding entry are provided by \citet{Higgins-et-al:1985, Gradstein:1995, Fu-Lu:2010,Kaplan-Sela:2010} and \citet{Fu-Jiao-Lu:2015}. Similarly, in the auction literature, endogenous entry has been modeled and analyzed by \citet{Levin-Smith:1994} and \citet{Pevnitskaya:2004}.\footnote{There is also a considerable amount of work that examines exogenous, or stochastic, entry into contests \citep[see, e.g.,][]{Munster:2006, Myerson-Warneryd:2006, Lim-Matros:2009, Fu-et-al:2011,Kahana-Klunover:2015, Kahana-Klunover:2016, Ryvkin-Drugov:2020} and into auctions \citep[see, e.g.,][]{Mcafee-Mcmillan:1987,Harstad-et-al:1990,Levin-Ozdenoren:2004}. For experimental evidence related to both exogenous and endogenous entry, see \citet{Anderson-Stafford:2003, Eriksson-et-al:2009, Morgan-et-al:2012, Morgan-et-al:2016, Hammond-et-al:2019, Boosey-et-al:2017, Boosey-Brookins-Ryvkin:2020, Aycinena-Rentschler:2019} in relation to contests, and \citet{Dyer-et-al:1989, Ivanova-Stenzel-Salmon:2004, Isaac-et-al:2012, Palfrey-Pevnitskaya:2008, Aycinena-Rentschler:2018} in relation to auctions.} In addition, much of the recent research on contest design has focused on the impact or optimality of different information disclosure policies. For instance, \citet{Lim-Matros:2009} and \citet{Fu-et-al:2011} consider the effect of disclosing the number of actual participants in contests with stochastic entry. They show that the disclosure policy is irrelevant for expected total effort in Tullock contests. However, \citet{Fu-et-al:2011} further show that in contests with a more general ``ratio-form'' CSF, the optimal disclosure rule depends on the shape of the CSF's impact function. Recent work by \citet{Ryvkin-Drugov:2020} generalizes these results by showing that the effect of disclosure in a general tournament model depends on the curvature of the cost function of effort.\footnote{See also \citet{Fu-Lu-Zhang:2016}, who study a generalized Tullock contest with two players who are asymmetric in terms of both their values and their stochastic entry probabilities.} For an all-pay auction environment, \cite{Chen-Jiang-Knyazev:2017} show that compared with full concealment, disclosing the number of actual participants in an all-pay auction with private values decreases the expected total investment if and only if the participants' cost functions are concave.

In all of the aforementioned studies, disclosure relates exclusively to the \textit{number} of entrants in the contest. Yet, there are also several studies that explore the impact of disclosing participants' (initially private) valuations on expected total investment. For the standard single-item all-pay auction environment, \citet{Morath-Munster:2008} establish that expected total effort is lower under complete information (i.e., with disclosure) than under private information (i.e., without disclosure). Their result is generalized to a contest with multiple prizes by \citet{Fu-Jiao-Lu:2014}.\footnote{For two other studies that explore a slightly richer set of disclosure policies, see \citet{Lu-Ma-Wang:2018} and \citet{Serena:2022}.} \citet{Feng:2023} considers an all-pay auction with entry, but also restricts attention to one dimension of disclosure (valuations), while \citet{Zhang-Zhou:2016} employ a Bayesian persuasion approach to show that, in general, contest designers may benefit by adopting a policy of partial disclosure.

To conclude, the existing literature has a lot to say about the effects of endogenous entry and disclosure in individual contests, and the main contribution of this paper is our extension of the analysis of these phenomena to contests among groups. The only other study that we are aware of that considers disclosure of the number of entrants in group contests is by \citet{Boosey-et-al:2019}, who examine Tullock contests among groups with stochastic group sizes and players with a common (and publicly known) prize valuation. Similarly, the only other study of disclosure of types within a group is \cite{Barbieri-et-al:2019}. In this paper, we consider the effects of disclosing both the number of entrants \emph{and} their private valuations. The structure of competition among groups also allows us to examine within-group disclosure---a particular form of partial disclosure, in which individuals learn about the number and valuations of entrants in their own group, but not of those in other groups.

\section{Model setup}
\label{sec:model}
There are $n\geqslant 2$ groups, indexed by $i=1,\ldots,n$, with $m\geqslant 1$ players in each group, indexed by $ij=i1,\ldots,im$.\footnote{In Section \ref{sec:hetero_sizes}, we briefly discuss the effect of disclosure for two groups with different sizes.} The players are risk neutral expected payoff maximizers. Each player $ij$ is endowed with prize valuation $v_{ij}$, which is initially the player's private information, drawn independently from a commonly known distribution with interval support $V=[\underline{v},\overline{v}]\subseteq\mathds{R}_+$, absolutely continuous cdf $F(\cdot)$ and continuous, positive a.e. pdf $f(\cdot)$.  

The game consists of two stages. In the first stage, each player $ij$ decides whether or not to enter the group contest as a member of group $i$. The players who decide to stay out receive an outside option payoff $\omega\in\mathds{R}_+$. The entrants proceed to the second stage. Let $M_i\subseteq\{i1,\ldots,im\}$ denote the set of entrants in group $i$. In the second stage, depending on a disclosure condition, some information may be revealed to entrants, after which they choose their investment levels $x_{ij}\in\mathds{R}_+$. Group $i$'s aggregate investment is determined by the best-shot technology as $X_i=\max_{ij\in M_i}x_{ij}$.\footnote{The case of additive technology is briefly discussed in Section \ref{sec:additive}.} The group with the highest investment wins the contest, and all entrants in that group receive their valuations. Entrants in all other groups receive zero. Ties are broken randomly, but occur with probability zero in equilibrium. All entrants pay their investments. 

\paragraph{Information disclosure} We consider three information settings implementing different modes of information disclosure at the beginning of the second stage, before entrants make their investment decisions. 

\noindent (i) No disclosure (ND): Entrant $ij$ knows only $v_{ij}$ (observed prior to entry).

\noindent (ii) Within-group disclosure (WD): Entrant $ij$ observes $v_{ik}$ for all $ik\in M_i$.

\noindent (iii) Full disclosure (FD): Entrant $ij$ observes $v_{lk}$ for all $lk\in \cup_{l=1}^nM_l$.

In the ND setting, no new information is revealed between the stages, and the game effectively collapses into one stage. In the case of within-group disclosure (WD), entrants observe others' valuations within their own groups. Finally, in the FD setting all entrants' valuations become public information. 

Importantly, disclosure happens in a centralized manner, e.g., via an information system, and entrants have no opportunity to voluntarily disclose their own types.\footnote{\cite{Barbieri-et-al:2019} consider group contests with ``weak-link'' aggregation where such voluntary disclosure within groups is possible.} Moreover, given our approach to within-group coordination in WD and FD (discussed below), in settings (ii) and (iii) it is sufficient for each entrant to only observe the highest valuation in their group and the highest valuations in all groups, respectively, and it is not necessary to observe the numbers of entrants.

\paragraph{Analysis} We look for a symmetric cutoff equilibrium in which there is a valuation $v^*\in V$ such that a player with valuation $v$ enters the contest if and only if $v\in V^*=[v^*,\overline{v}]$.\footnote{If $v^*>\underline{v}$ then, by continuity, the marginal player with $v=v^*$ is indifferent between entering and staying out. For concreteness, and without loss, we assume that such a player enters.} We assume that $\omega<\overline{v}$ so that at least some entry occurs with positive probability. Let $q=1-F(v^*)$ denote the \textit{ex ante} probability of entry. Further, let $\tilde{F}(v)=\frac{F(v)-F(v^*)}{q}\mathds{1}_{v\geqslant v^*}$ and $\tilde{f}(v)=\frac{f(v)}{q}\mathds{1}_{v\in V^*}$ denote the updated cdf and pdf of entrants' valuations.

\paragraph{The principal's objective} 

Throughout our main analysis, we (implicitly) consider a principal whose objective is to maximize the \emph{expected aggregate investment} of all groups, $\mathds{E}(\sum_{i=1}^nX_i)$. This objective is standard in the literature and suitable for an organizational setting where the manager who oversees a portfolio of departments or teams cares about the investments from all of them.\footnote{Consider, for example, a large tech firm, where the R\&D team may be split into several independent units or divisions, each tasked with proposing new features for the firm's product line. In such an organization, the manager generally cares about the quality of the best proposals developed by all of the units, even if only the overall best is rewarded according to the contest rules.} Note that the principal's objective (aggregate of all best-shot investments) and the contest prize structure (overall best-shot only) need not coincide.\footnote{This is apparent in the individual contests literature where a winner-takes-all contest is often optimal to incentivize the sum of efforts.} Nevertheless, there are also settings in which the principal only cares about the best overall output (as in a pure winner-takes-all innovation contest), and settings in which the principal cares about all individuals' investments, including those that are dominated by others within their teams. In Section \ref{sec:other_objectives}, we extend our analysis to consider these two alternative objectives---\emph{expected total investment}, $\mathds{E}(\sum_{i=1}^n\sum_{ij\in M_i}x_{ij})$, and \emph{expected highest investment}, $\mathds{E}(\max_{i\in\{1,\ldots,n\}}X_i)$---while maintaining the (overall) best-shot prize structure.

\section{Contests among individuals ($m=1$)}
\label{sec:individuals}
As an important benchmark setting, we first consider contests among individuals. Within-group disclosure always takes place in this case, by definition, and is equivalent to no disclosure; therefore, we only compare no disclosure and full disclosure. Observe that the cutoff valuation, $v^*$, is independent of disclosure. Indeed, without disclosure the marginal player will invest zero and can only win the contest if she is the only entrant. In all other cases, she will lose with probability one and earn zero payoff. Under (full) disclosure, the marginal player will also invest zero and win if she is the only entrant. If there are several entrants, the investment stage game is an all-pay auction of complete information, with equilibrium in mixed strategies \citep[see, e.g.,][]{Baye-et-al:1996}. However, the marginal player's valuation will be the lowest with probability one; therefore, in equilibrium she will earn zero in expectation.

Thus, the cutoff type, $v^*$, is determined by the indifference condition equating the payoff of the marginal player when she is the only entrant to the outside option:
\begin{equation}
\label{cutoff_ind}
v^*F(v^*)^{n-1} = \omega.
\end{equation}
Under our assumption that $\omega<\overline{v}$, Eq. (\ref{cutoff_ind}) has a unique solution.

\paragraph{No disclosure} Let $b^{(1)}_{\rm N}(v)$ denote the symmetric monotone bidding function for entrants (here and in what follows, subscript ``N'' stands for nondisclosure; superscript ``(1)'' distinguishes contests between individuals). Using the standard approach, consider an entrant with valuation $v\in V^*$ bidding as if her valuation is $\hat{v}\in V^*$. The entrant's payoff is then given by $\Pi^{(1)}(v,\hat{v};v^*) = vF(\hat{v})^{n-1}-b^{(1)}_{\rm N}(\hat{v})$, where $F(\hat{v})^{n-1}$ is the probability that her bid is the highest. The first-order condition $\Pi^{(1)}_{\hat{v}}(v,v;v^*)=0$ produces a differential equation for the unknown bidding function $b^{(1)}_{\rm N}(v)$, which, together with the initial condition $b^{(1)}_{\rm N}(v^*)=0$, gives
\begin{equation}
\label{b_nd_ind}
b^{(1)}_{\rm N}(v) = (n-1)\int_{v^*}^v tF(t)^{n-2}\dd F(t).
\end{equation}
Following the standard argument, we verify that $\Pi^{(1)}_{\hat{v}}(v,\hat{v};v^*)$ changes sign at $\hat{v}=v$ in a way that makes $b^{(1)}_{\rm N}(v)$ given by (\ref{b_nd_ind}) the unique symmetric equilibrium bidding function for a given $v^*$. Equation (\ref{b_nd_ind}) has exactly the same form as the equilibrium bidding function in independent private value all-pay auctions without entry \citep[see, e.g.,][]{Krishna-Morgan:1997}, except it is shifted down by a constant and truncated at $v^*$. The effect of endogenous entry is contained entirely in $v^*$.

Expected aggregate investment under no disclosure, therefore, is
\begin{align}
\label{B_nd_ind}
B^{(1)}_{\rm N} = nq\int_{v^*}^{\overline{v}}b^{(1)}_{\rm N}(v)\dd\tilde{F}(v) 
= n(n-1)\int_{t_1\geqslant t_2\geqslant v^*}t_2F(t_2)^{n-2}\dd F(t_1)\dd F(t_2).
\end{align}

\paragraph{Full disclosure} Suppose there are $k\geqslant 2$ entrants,\footnote{States with one or zero entrants contribute nothing to aggregate investment.} and let $v_{(1)}>v_{(2)}\ldots>v_{(k)}$ denote their ranked valuations (ties in valuations are probability zero events; therefore, we can generically assume the strict inequalities). In equilibrium, the entrants with valuations $v_{(1)}$ and $v_{(2)}$ bid according to mixed strategies with common support $[0,v_{(2)}]$ and cdfs $G_1(x_1) = \frac{x_1}{v_{(2)}}$ and $G_2(x_2) = 1-\frac{v_{(2)}}{v_{(1)}}+\frac{x_2}{v_{(1)}}$, respectively, while all other entrants bid zero. The corresponding probabilities of winning are $p_1=1-\frac{v_{(2)}}{2v_{(1)}}$ and $p_2=\frac{v_{(2)}}{2v_{(1)}}$. Average bids are $b_1 = \frac{v_{(2)}}{2}$ and $b_2 = \frac{v_{(2)}^2}{2v_{(1)}}$. Finally, the expected payoffs are $\pi_1 = v_{(1)}-v_{(2)}$ and $\pi_2=0$, respectively \citep{Baye-et-al:1996}.

Expected aggregate investment under full disclosure, therefore, is
\begin{align*}
B^{(1)}_{\rm F} = \sum_{k=2}^n\binom{n}{k}q^k(1-q)^{n-k}\int_{t_1\geqslant t_2\geqslant v^*}\left(\frac{t_2}{2}+\frac{t_2^2}{2t_1}\right)\tilde{f}_{(1,2:k)}(t_1,t_2)\dd t_1\dd t_2,
\end{align*}
where $\tilde{f}_{(1,2:k)}(t_1,t_2)$ is the joint pdf of the top two order statistics in a sample of size $k$ from distribution $\tilde{F}(\cdot)$ (subscript ``F'' stands for full disclosure). This pdf is given by \citep{David-Nagaraja:2003}
\[
\tilde{f}_{(1,2:k)}(t_1,t_2) = k(k-1)\tilde{F}(t_2)^{k-2}\tilde{f}(t_1)\tilde{f}(t_2)\mathds{1}_{t_1\geqslant t_2},
\]
resulting in
\begin{align}
\label{B_fd_ind}
B^{(1)}_{\rm F} = n(n-1)\int_{t_1\geqslant t_2\geqslant v^*}\left(\frac{t_2}{2}+\frac{t_2^2}{2t_1}\right)F(t_2)^{n-2}\dd F(t_1)\dd F(t_2).
\end{align}
Comparing (\ref{B_nd_ind}) and (\ref{B_fd_ind}), we observe that $B^{(1)}_{\rm F}<B^{(1)}_{\rm N}$. This gives our first result.
\begin{proposition}
\label{prop_ind}
In contests among individuals, expected aggregate investment under (full) disclosure is lower than under no disclosure: $B^{(1)}_{\rm F}<B^{(1)}_{\rm N}$.
\end{proposition}

Proposition \ref{prop_ind} is an important benchmark result that serves as a motivation for what follows. It shows that in contests among individuals the disclosure of types has an unambiguous negative effect on aggregate investment. It holds for any cutoff valuation $v^*$, which includes contests where the number of players is fixed ($v^*=\underline{v}$), generalizing the result of \cite{Morath-Munster:2008}. It also holds when the number of players is stochastic following an exogenous distribution.

The mechanism behind Proposition \ref{prop_ind} is as follows. Without disclosure, the allocation of the prize is efficient and aggregate investment is given by the expectation of the second highest valuation, $v_{(2)}$---the same as in other revenue-equivalent auctions, such as the first-price or second-price auction. The expectation is truncated due to endogenous entry, but since the cutoff type is independent of disclosure, this truncation is irrelevant. Under disclosure, the efficiency is lost; \emph{and} the player with valuation $v_{(2)}$ bids zero with probability $1-\frac{v_{(2)}}{v_{(1)}}$, and hence her contribution to aggregate investment is reduced ($\frac{v^2_{(2)}}{2v_{(1)}}<\frac{v_{(2)}}{2}$ with probability one).
\section{Contests among groups ($m\geqslant 2$)}
\label{sec:groups}

In this section, we consider contests among groups of at least two potential players ($m\geqslant 2$), starting with the no disclosure (ND) case in Section \ref{sec:nd}. In Section \ref{sec:wd}, we then consider within-group disclosure (WD) and compare it to ND. In Section \ref{sec:fd}, we characterize full disclosure (FD) and compare it to both ND and WD.

\subsection{No disclosure}
\label{sec:nd}
Let $b_{\rm N}(v)$ denote the symmetric monotone bidding function for entrants. Again following the standard approach, consider an entrant with valuation $v\in V^*$ bidding according to a valuation $\hat{v}\in V^*$. The entrant's payoff is $\Pi(v,\hat{v};v^*) = vp_{\rm N}(\hat{v})-b_{\rm N}(\hat{v})$, where the probability of winning is
\begin{align}
\label{p_nd}
&p_{\rm N}(\hat{v}) = (1-q)^{nm-m}+\sum_{k_1=0}^{m-1}\sum_{k_2=1}^{nm-m}\binom{m-1}{k_1}\binom{nm-m}{k_2}q^{k_1+k_2}(1-q)^{nm-1-k_1-k_2}\times\nonumber\\
&\times\left[\tilde{F}(\hat{v})^{k_1+k_2}+(1-\tilde{F}(\hat{v})^{k_1+k_2})\frac{k_1}{k_1+k_2}\right].
\end{align}
The first term represents the situation when there are no entrants in other groups. The second term sums over all possible configurations of the numbers of entrants in the player's own group ($k_1$) and other groups ($k_2$). The first term in square brackets is the probability that the player's valuation (and hence the bid) is the highest of them all, while the second term is the ``free-riding component'' where the player's valuation is not the highest but she nevertheless wins because someone else in her group has the highest valuation.

The first-order condition $\Pi_{\hat{v}}(v,v;v^*)=0$ produces a differential equation for the unknown bidding function $b_{\rm N}(v)$, which, together with the initial condition $b_{\rm N}(v^*)=0$, gives
\begin{equation}
\label{b_nd}
b_{\rm N}(v) = m(n-1)\int_{v^*}^v tF(t)^{nm-2}\dd F(t).
\end{equation}
The details of the derivation are provided in Appendix \ref{sec:appendix_Eq6}. We again verify that $\Pi_{\hat{v}}(v,\hat{v};v^*)$ changes sign at $\hat{v}=v$ in a way that makes $b_{\rm N}(v)$ given by (\ref{b_nd}) the unique symmetric equilibrium bidding function for a given $v^*$. Similar to individual contests, Eq. (\ref{b_nd}) has exactly the same form as the equilibrium bidding function in a group contest with a fixed number of players \citep{Barbieri-Malueg:2016}, except it is shifted down by a constant and truncated at $v^*$.

In order to identify $v^*$, suppose $v^*\in{\rm int}(V)$ and hence the cutoff type is indifferent between entering and not entering. Her payoff from entry is $\Pi(v^*,v^*;v^*) = v^*p_{\rm N}(v^*)$, where
\begin{align}
\label{p_nd_marginal}
&p_{\rm N}(v^*) = (1-q)^{nm-m}+\sum_{k_1=0}^{m-1}\sum_{k_2=1}^{nm-m}\binom{m-1}{k_1}\binom{nm-m}{k_2}q^{k_1+k_2}(1-q)^{nm-1-k_1-k_2}\frac{k_1}{k_1+k_2}\nonumber\\
& = \frac{m-1}{nm-1} + \frac{m(n-1)}{nm-1}F(v^*)^{nm-1}.
\end{align}
For details, see Appendix \ref{sec:appendix_Eq7}. It follows that $v^*=\underline{v}$, i.e., there is full entry, if $\omega\leqslant  \frac{m-1}{nm-1}\underline{v}$. Otherwise, the cutoff is given by the unique solution of the equation $v^*p_{\rm N}(v^*)=\omega$.\footnote{As seen from (\ref{p_nd_marginal}), $p_{\rm N}(v)$ is continuous and (weakly) increasing in $V$ and hence $vp_{\rm N}(v)$ is continuous and strictly increasing.}

Expected aggregate investment in each group is given by the expectation of the maximum bid among entrants, which, together with (\ref{b_nd}), gives expected aggregate investment in the contest:
\begin{align}
\label{B_nd}
&B_{\rm N} = n\sum_{k=1}^m\binom{m}{k}q^k(1-q)^{m-k}\int_{v^*}^{\overline{v}}b_{\rm N}(v)\dd\tilde{F}(v)^k = n\int_{v^*}^{\overline{v}}b_{\rm N}(v)\dd F(v)^m\nonumber\\
& = m^2n(n-1)\int_{t_1\geqslant t_2\geqslant v^*}t_2F(t_2)^{nm-2}F(t_1)^{m-1}\dd F(t_1)\dd F(t_2).
\end{align}

\subsection{Within-group disclosure}
\label{sec:wd}

In this setting, valuations of entrants are revealed within each group before the group members decide on their investments. From an organizational design perspective, a principal might enforce this WD structure---obscuring information between competing units while encouraging internal transparency, specifically to facilitate  coordination.

While it is clear that only one of the entrants within each group will be active in any (pure strategy) equilibrium, multiple such equilibria are possible, with different entrants being active. We assume that groups coordinate so that the \textit{leaders}---the entrants with the highest valuations---are the active bidders. This assumption is quite reasonable in the case of best-shot aggregation where the most capable group member is the natural leader. It is also the only equilibrium that is parallel to the one arising in the setting without disclosure where each group's bid is by construction determined by the highest-valuation entrant.\footnote{As an alternative justification, suppose each group has a non-bidding manager who derives some value from the group winning the contest and is able to select one active bidder among the entrants in her group (subject to the entrants' participation constraints). Then each manager would select the entrant in her group with the highest valuation to be the active bidder. Similarly, this outcome would emerge if all entrants could select the active bidder via a binding majority voting procedure (with the highest valuation entrant appointed in the case of a tie when there are two entrants).} 

We again look for a symmetric cutoff entry equilibrium with some marginal type $v^*$. Let $b_{\rm W}(v)$ denote the monotone bidding function of leaders in each group (subscript ``W'' stands for within-group disclosure). Following the same steps as in Section \ref{sec:nd}, consider a leader with valuation $v\in V^*$ that is bidding as if her valuation is $\hat{v}\in V^*$. The leader's payoff is $\Pi(v,\hat{v};v^*)=vF(\hat{v})^{nm-m}-b_{\rm W}(\hat{v})$. Indeed, this leader's group wins if her bid exceeds that of all leaders or, equivalently, of all potential players in other groups. Solving the first-order condition $\Pi_{\hat{v}}(v,v;v^*)=0$, we obtain the bidding function
\begin{equation}
\label{b_wd}
b_{\rm W}(v) = m(n-1)\int_{v^*}^{v}tF(t)^{nm-m-1}\dd F(t).
\end{equation}
The optimality of $b_{\rm W}(v)$ follows similar to Section \ref{sec:nd}. All entrants whose valuations are not the highest bid zero.\footnote{We verify in Appendix \ref{sec:appendix_Eq9} that none of the entrants can deviate profitably by bidding a positive amount.}

Consider now the payoff of the marginal type $v^*$ in the case of entry. This type always bids zero and can win if there are no entrants in other groups or the leader in her group is the winner. Thus, the payoff of the marginal entrant is exactly the same as in the no disclosure case, and hence the equilibrium $v^*$ under within-group disclosure is the same as the $v^*$ under no disclosure identified in Section \ref{sec:nd}.

From (\ref{b_wd}), the expected aggregate investment by entrants is
\begin{align}
\label{B_wd}
B_{\rm W} = n\sum_{k=1}^m\binom{m}{k}q^k(1-q)^{m-k}\int_{v^*}^{\overline{v}}b_{\rm W}(v)\dd\tilde{F}(v)^k = n\int_{v^*}^{\overline{v}}b_{\rm W}(v)\dd F(v)^m\nonumber\\
=m^2n(n-1)\int_{t_1\geqslant t_2\geqslant v^*}t_2F(t_2)^{nm-m-1}F(t_1)^{m-1}\dd F(t_1)\dd F(t_2).
\end{align}
Using Eqs. (\ref{B_nd}) and (\ref{B_wd}), the difference in expected aggregate investment between the within-group disclosure and no disclosure settings is\footnote{Alternatively, observe from Eqs. (\ref{b_nd}) and (\ref{b_wd}) that $b_{\rm W}(v)\geqslant b_{\rm N}(v)$ for each $v$, i.e., the two individual bidding functions are clearly ranked pointwise. Since aggregate group investment comes from the highest valuation group member in both cases, the comparison between $B_{\rm N}$ and $B_{\rm W}$ follows immediately. We rely on this approach when we discuss the ranking of the expected highest investment in Section \ref{sec:other_objectives}, but here we compute $B_{\rm N}$ and $B_{\rm W}$ directly to facilitate the comparisons with full disclosure in Section \ref{sec:fd}.}
\begin{align*}
&B_{\rm W}-B_{\rm N}\\
& = m^2n(n-1)\int_{t_1\geqslant t_2\geqslant v^*}t_2F(t_2)^{nm-m-1}F(t_1)^{m-1}\left[1-F(t_2)^{m-1}\right]\dd F(t_1)\dd F(t_2)\geqslant 0.
\end{align*}
Note that the inequality above is strict whenever $m>1$, i.e., when within-group disclosure can actually reveal new information. Thus, we arrive at the following result. 

\begin{proposition}
\label{prop_comp_nd_wd}
Expected aggregate investment under within-group disclosure is greater than under no disclosure, $B_{\rm W}>B_{\rm N}$.
\end{proposition}

The difference stems from the fact that the leaders' identities in the no disclosure case are unknown; every entrant can be a leader with some probability, and there is always some probability that an entrant's bid will be wasted. With disclosure, the entrants are able to solve the coordination problem within groups and bid more effectively.

While this result pertains to a principal concerned with expected aggregate investment, we show in Section~\ref{sec:other_objectives} that the effect of within-group disclosure on the leader's investment is so strong that it strictly exceeds also the expected sum of the (heavily shaded) investments by non-leaders under ND. That is, while WD dominates ND according to our main objective criterion (expected aggregate investment), it also dominates ND according to the alternative objective of maximizing the expected \textit{total} investment.

\subsection{Full disclosure}
\label{sec:fd}
In this setting, valuations of all entrants are revealed across groups. Similar to Section \ref{sec:wd}, we assume that within each group players coordinate so that the groups' leader---the highest-valuation entrant---is its (potentially) active bidder. In this case, the second stage effectively turns into an all-pay auction of complete information among the leaders. If there are entrants in two or more groups, the unique equilibrium (with probability one) involves two active group leaders with the highest valuations bidding according to mixed strategies while all other groups drop out.

We again look for a cutoff entry equilibrium with some marginal type $v^*$. The payoff of the marginal type from entry is $v^*p_{\rm F}(v^*)$, where 
\begin{align}
\label{p_fd_m}
& p_{\rm F}(v^*) = (1-q)^{nm-m} + (n-1)\sum_{k_1=0}^{m-1}\sum_{k_2=1}^m\sum_{k_3=0}^{nm-2m}\binom{m-1}{k_1}\binom{m}{k_2}\binom{nm-2m}{k_3}\times\nonumber\\
&\times q^{k_1+k_2+k_3}(1-q)^{nm-1-k_1-k_2-k_3}\left[\int_{t_1\geqslant t_2\geqslant t_3\geqslant v^*}\left(1-\frac{t_2}{2t_1}\right)\dd\tilde{F}(t_1)^{k_1}\dd\tilde{F}(t_2)^{k_2}\dd\tilde{F}(t_3)^{k_3} + \right.\nonumber\\
&\left. + \int_{t_2\geqslant t_1\geqslant t_3\geqslant v^*}\frac{t_1}{2t_2}\dd\tilde{F}(t_1)^{k_1}\dd\tilde{F}(t_2)^{k_2}\dd\tilde{F}(t_3)^{k_3} \right].  
\end{align}
As before, the first term describes the case when there are no entrants in other groups. The triple summation goes over the possible numbers of other entrants in the marginal entrant's own group ($k_1$), entrants in a second group ($k_2$), and entrants in all other groups ($k_3$). There are $n-1$ possible second groups in this context, hence the multiplier $(n-1)$. Winning occurs with positive probability when the marginal entrant's own group has a leader who is among the top two leaders; hence, integration is restricted to the domain with $\min\{t_1,t_2\}\geqslant t_3$. The equilibrium probabilities of winning (cf. Section \ref{sec:individuals}) are used for the cases with $t_1\geqslant t_2$ and $t_2\geqslant t_1$ in the two integrals. Integration is over the highest order statistics in all cases.\footnote{Note that, strictly speaking, $p_{\rm F}$ is not the \textit{probability} of winning. The marginal entrant can also win with positive probability if she is her group's leader (i.e., the only entrant) and there is only one other group with entrants. However, in this case the marginal entrant's expected payoff is zero, cf. Section \ref{sec:individuals}.}

Simplifying (\ref{p_fd_m}), we obtain $p_{\rm F}(v^*) = p_{\rm N}(v^*) + A(v^*)$, where $p_{\rm N}(v^*)$ is the marginal type's probability of winning in the case of nondisclosure, Eq. (\ref{p_nd}), and
\begin{align}
\label{p_fd_marginal}
&A(v^*) = m(m-1)(n-1)\int_{t_1\geqslant t_2\geqslant v^*}\frac{t_2}{2t_1}F(t_1)^{m-2}F(t_2)^{nm-m-2}\times\nonumber\\
&\times [F(t_1)-F(t_2)]\dd F(t_1)\dd F(t_2)\geqslant 0.
\end{align}
For details, see Appendix \ref{sec:appendix_Eq12}. Thus, the marginal type $v^*_{\rm F}$ such that $v^*_{\rm F}p_{\rm F}(v^*_{\rm F})=\omega$ is lower than in the case of no disclosure, $v^*_{\rm F}\leqslant v^*$, and the inequality is strict for $m>1$ and $v^*>\underline{v}$. A larger mass of players enters the contest under full disclosure.

\begin{proposition}
\label{prop_cutoffs}
Suppose there is less than full entry under no disclosure, $v^*>\underline{v}$. Then there is more entry under full disclosure than under no disclosure: $v^*_{\rm F}<v^*$.
\end{proposition}

Intuitively, a marginal entrant can win the contest in ND (and in WD) only if her group's leader has the highest valuation among all entrants in all groups. However, in FD it is also possible to win the contest with positive probability if the group leader has the second highest valuation. This probability, for a given realization of valuations, is $\frac{t_2}{2t_1}$, which is the term appropriately averaged in $A(v^*)$. It raises the payoff of the marginal entrant, and hence the cutoff type is lower in FD as compared to the other two cases.

Given the marginal type $v^*_{\rm F}$, it can be shown that the expected payoff from entering is less than $\omega$ for all types $v < v_{\rm F}^*$, and greater than $\omega$ for all types $v > v_{\rm F}^*$. First, for $v < v_{\rm F}^*$, type $v$ either earns $v$ or zero, and her probability of earning zero is the same as for the marginal type. Thus, the expected payoff of type $v < v_{\rm F}^*$ from entering must be lower than the expected payoff of the marginal type, which is, by definition, equal to $\omega$. Second, for $v \geqslant v_{\rm F}^*$, we show in Appendix \ref{sec:appendix_payoff_increasingFD} that the expected payoff from entering, $\Pi(v,v;v_{\rm F}^*)$, is increasing in $v$.

Recall our assumption that $\omega<\overline{v}$. The existence of a marginal type (i.e., an interior solution to $v^*_{\rm F}p_{\rm F}(v^*_{\rm F})=\omega$) is then guaranteed by the following sufficient condition:
\begin{align}
\label{exist_fd_sufficient}
&\frac{\omega}{\underline{v}} \geqslant \frac{m-1}{nm-1} \left(1 + \frac{m(n-1)}{2(nm-1)(nm-m-1)} \right).
\end{align} 

The following example illustrates the comparison of equilibrium cutoffs for the full disclosure and no disclosure settings (Proposition \ref{prop_cutoffs}).

\begin{figure}[!t]
\begin{center}
\subfloat{
\epsfig{figure=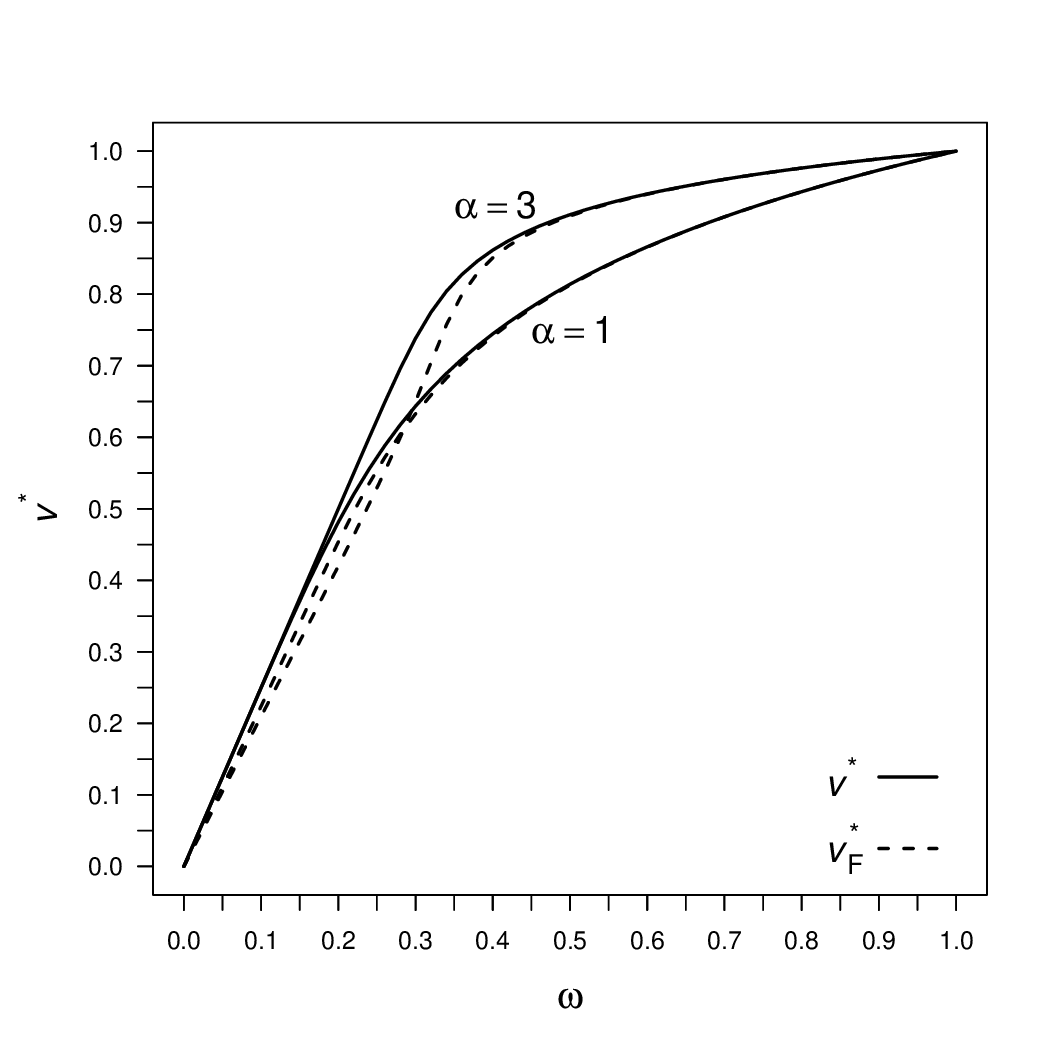,width=3in}
\label{cutoff}}
\subfloat{
\epsfig{figure=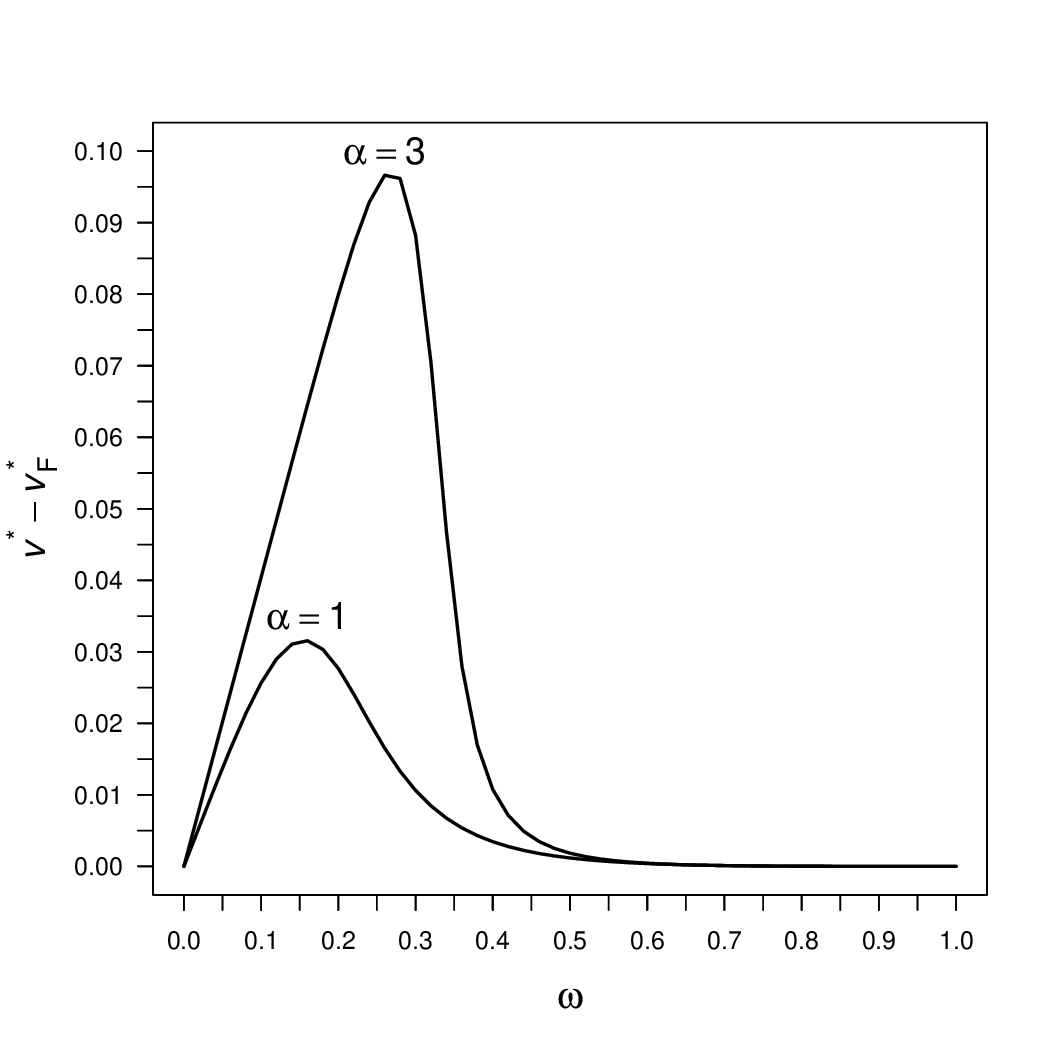,width=3in}
\label{diff}}
\caption{Cutoff valuations for ND and FD as a function of $\omega$ (left), and the difference in cutoff valuations, $v^{*} - v_{\rm F}^{*}$, as a function of $\omega$ (right), in Example \ref{example1}.}
\label{fig:example1}
\end{center}
\end{figure}

\begin{example}
\label{example1}
Suppose $n = 2$, $m = 3$, and that valuations are drawn from a distribution of the form $F(v) = v^{\alpha}$, $\alpha>0$,  with interval support $V = [0,1]$. In Figure \ref{fig:example1}, we plot the equilibrium cutoff valuation (type) as a function of the outside option, $\omega$, for both the ND and FD cases with $\alpha = 1$ (uniform distribution) and $\alpha = 3$.
\end{example}

From the left panel of Figure \ref{fig:example1}, which plots the cutoff valuations, it can be difficult to see that $v_{\rm F}^* < v^*$ for all (interior) values of $\omega$. Thus, in the right panel, we also plot the difference in cutoff valuations, $v^{*} - v_{\rm F}^{*}$, as a function of $\omega$, for the two different values of $\alpha$. This better illustrates that the cutoff type is higher under ND than under FD, but also serves to illustrate that the difference is non-monotone and single-peaked. The nonmonotonicity of the difference $v^{*} - v_{\rm F}^{*}$ is expected because the two cutoffs have to be the same under full entry and no entry. 

Expected aggregate investment under full disclosure is
\begin{align}
\label{B_fd_def}
& B_{\rm F} = n(n-1)\sum_{k_1=1}^m\sum_{k_2=1}^m\binom{m}{k_1}\binom{m}{k_2}q^{k_1+k_2}(1-q)^{2m-k_1-k_2}\int_{t_1\geqslant t_2\geqslant v^*_{\rm F}}\left(\frac{t_2}{2}+\frac{t_2^2}{2t_1}\right) \nonumber\\
& \times\sum_{k_3=0}^{nm-2m}\binom{nm-2m}{k_3}q^{k_3}(1-q)^{nm-2m-k_3}\tilde{F}(t_2)^{k_3}\dd\tilde{F}(t_1)^{k_1}\dd\tilde{F}(t_2)^{k_2}.
\end{align}
Here, as above, we select two groups with the two highest leaders' valuations, $t_1$ and $t_2$, assuming $t_1\geqslant t_2$. To generate positive aggregate investment, there must be at least one entrant in each of these groups. The remaining groups may have any number of entrants, $k_3$, whose valuations are all below $t_2$. There are $n(n-1)$ ways to pick the initial two groups.

Simplifying, we obtain
\begin{equation}
\label{B_fd}
B_{\rm F} = m^2n(n-1)\int_{t_1\geqslant t_2\geqslant v^*_{\rm F}}\left(\frac{t_2}{2}+\frac{t_2^2}{2t_1}\right)F(t_1)^{m-1}F(t_2)^{nm-m-1}\dd F(t_1)\dd F(t_2).
\end{equation}
For details, see Appendix \ref{sec:appendix_Eq15}. The following proposition (also proved in Appendix \ref{sec:appendix_Eq15}) is our second major result.

\begin{proposition}
\label{prop_comp_nd_fd}
Suppose the elasticity of the distribution of types, $\xi(t)=\frac{tf(t)}{F(t)}$, satisfies $\xi(t)\geqslant \frac{1}{m-1}$. Then expected aggregate investment under full disclosure is greater than under no disclosure, $B_{\rm F}>B_{\rm N}$.
\end{proposition}

The proof of Proposition \ref{prop_comp_nd_fd} is based on comparing $B_{\rm F}$, Eq. (\ref{B_fd}), to $B_{\rm N}$, Eq. (\ref{B_nd}). Recall that $v^*_{\rm F}\leqslant v^*$; thus, the domain of integration for $B_{\rm F}$ is larger. The lower bound on the elasticity of $F(\cdot)$ ensures that the integrand in (\ref{B_fd}) is also larger.

The next example illustrates the comparison of aggregate investment between FD and ND when the sufficient condition on the elasticity of the distribution of types provided by Proposition \ref{prop_comp_nd_fd} is not satisfied. In particular, it demonstrates that the ranking of aggregate investment may be (but need not be) reversed.

\begin{figure}[!t]
\begin{center}
\subfloat{
\epsfig{figure=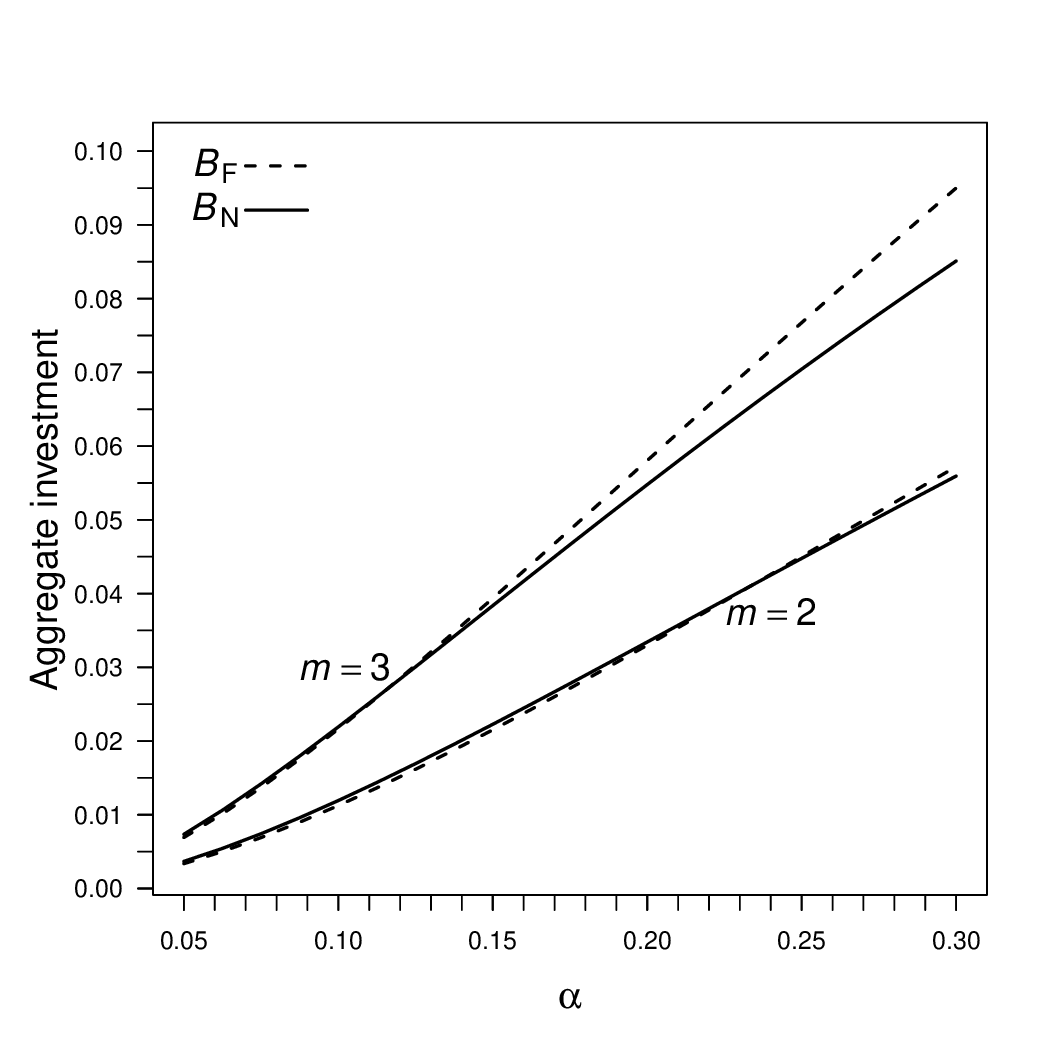,width=3in}
\label{agginvest}}
\subfloat{
\epsfig{figure=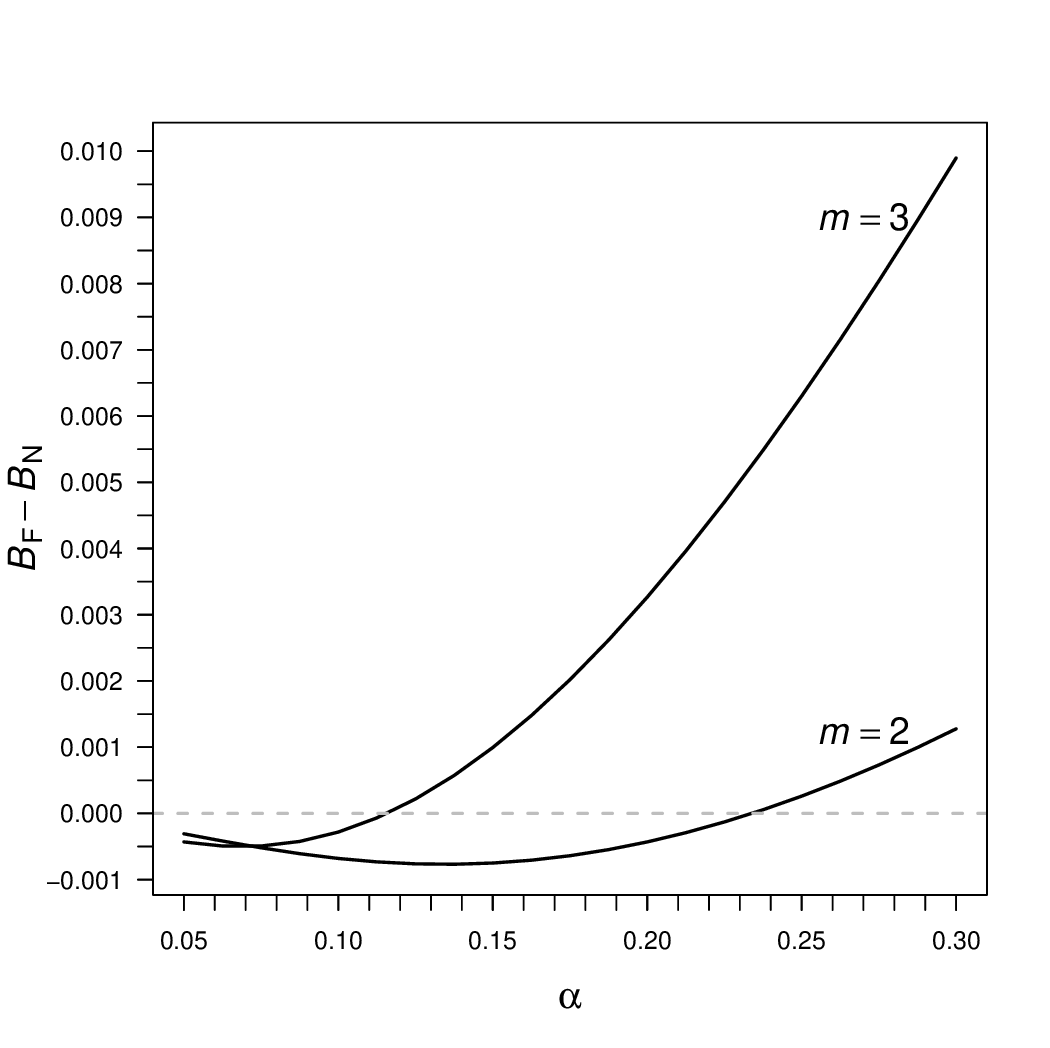,width=3in}
\label{agginvestdiff}}
\caption{Aggregate investment for ND and FD as a function of $\alpha$, for $m=2$ and $m=3$ (left), and the difference, $B_{\rm F}-B_{\rm N}$, as a function of $\alpha$, for $m=2$ and $m=3$ (right), Example \ref{example2}.}
\label{fig:example2}
\end{center}
\end{figure}

\begin{example}
\label{example2}
As in Example \ref{example1}, suppose $n = 2$, $V = [0,1]$, and $F(v) = v^{\alpha}$. Fix $\omega = 0.4$ and consider two cases corresponding to $m=2$ and $m = 3$. In Figure \ref{fig:example2}, we plot the aggregate investment levels for values of $\alpha$ (the elasticity of the distribution of types) between 0.05 and 0.3. 
\end{example}

Proposition \ref{prop_comp_nd_fd} implies that, for $m = 2$ (respectively, $m = 3$), $\alpha \geqslant 1$ (respectively, $\alpha \geqslant 0.5$) is sufficient for $B_{\rm F} > B_{\rm N}$. Thus, the sufficient condition is not satisfied for either $m$ in Figure \ref{fig:example2}.

The left panel establishes that it is possible for aggregate investment to be lower under FD than under ND. For both values of $m$, $B_{\rm F}$ (the dashed lines) is below $B_{\rm N}$ (the solid lines) for at least some values of $\alpha$. In order to better highlight the comparison, we also plot the difference, $B_{\rm F} - B_{\rm N}$, as a function of $\alpha$, in the right panel.

When $m = 3$, $B_{\rm F} < B_{\rm N}$ for very small values of $\alpha$. However, as the elasticity of the distribution of types increases, aggregate investment under FD grows much faster than under ND such that, even though the condition of Proposition \ref{prop_comp_nd_fd} is not satisfied, we still obtain $B_{\rm F} > B_{\rm N}$. Similarly, when $m = 2$, if $\alpha$ is low, aggregate investment is higher under ND than under FD. Still, once $\alpha$ becomes large enough, full disclosure leads to higher aggregate investment than no disclosure.

The intuition for this relationship between the elasticity of the distribution of types and the effect of full disclosure is as follows. The effect of full disclosure on aggregate investment depends on three competing effects, two of which lead to an increase and one---to a decrease in aggregate investment as compared to ND. First, as is the case for the WD setting, full disclosure allows entrants to solve the coordination problem within their own group, which has a positive effect on expected group-level investment, as it reduces the effect of free riding on the highest valuation entrant's investment. 

Second, full disclosure reduces the number of active bidders among entrants to just two (the two highest valuation leaders of their respective groups) who play according to the standard  mixed-strategy equilibrium in an all-pay auction of complete information. In this equilibrium, the second-highest valuation player places a mass on zero investment that reduces the expected aggregate investment.\footnote{The equilibrium is also inefficient, as the lower valuation bidder wins with positive probability. Notably, this negative effect of full disclosure on aggregate investment is \emph{the only} effect that is present in contests among individuals, which explains the contrast between Propositions \ref{prop_comp_nd_fd} and \ref{prop_ind}.} Importantly, this mass is increasing in the difference between the highest and second-highest valuations. As the elasticity of $F(\cdot)$ increases, draws from the distribution of valuations shift closer to the upper bound of the support, such that, in expectation, the difference between the highest and second-highest valuations becomes smaller. Thus, with an increase in the elasticity of the distribution, the negative effect of full disclosure becomes less important, allowing for FD to increase aggregate investment above the level in the ND setting. We note that the role of elasticity of the distribution of types in comparative statics for group contests has also been identified by \cite{Barbieri-Malueg:2016}, who showed that it affects how aggregate investment changes with group size.

Finally, due to Proposition \ref{prop_cutoffs} there is more entry under FD, which also increases aggregate investment. This explains why the sufficient condition in Proposition \ref{prop_comp_nd_fd} is not very tight, cf. Example \ref{example2}. 

Note that a similar comparison cannot be made between $B_{\rm F}$ and $B_{\rm W}$, Eq. (\ref{B_wd}). While the domain of integration is larger in $B_{\rm F}$, we observe that the integrand is always larger in $B_{\rm W}$. These competing effects---a larger mass of players entering but bidding lower under full disclosure---make the comparison ambiguous. As such, there is no systematic condition, independent of the cutoff, that suffices to establish an unambiguous ranking of aggregate investment between WD and FD. Our third example demonstrates this ambiguity, and shows that even when the condition in Proposition \ref{prop_comp_nd_fd} is satisfied, it is possible for $B_{\rm F}$ to be higher or lower than $B_{\rm W}$. 

\begin{figure}[!t]
\begin{center}
\subfloat{
\epsfig{figure=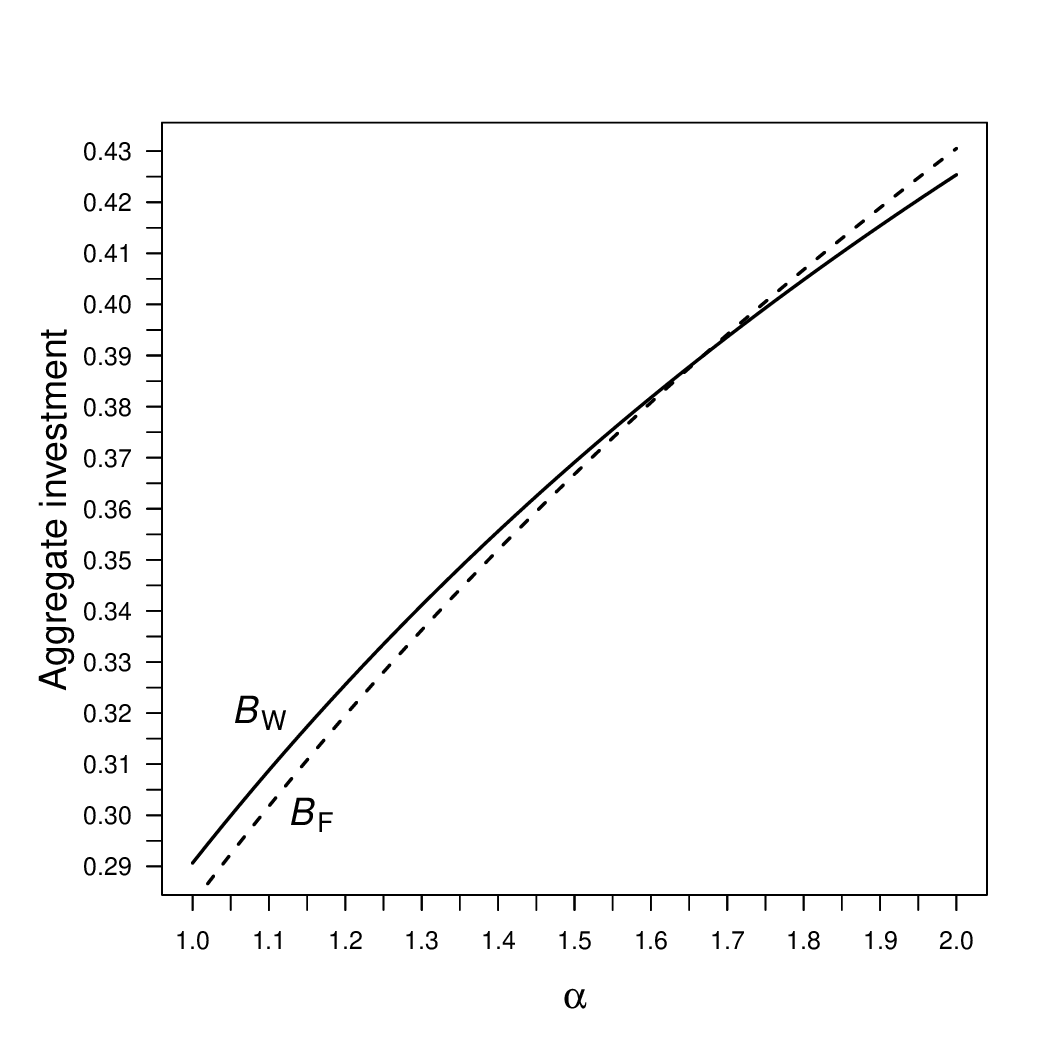,width=3in}
\label{ex3a}}
\subfloat{
\epsfig{figure=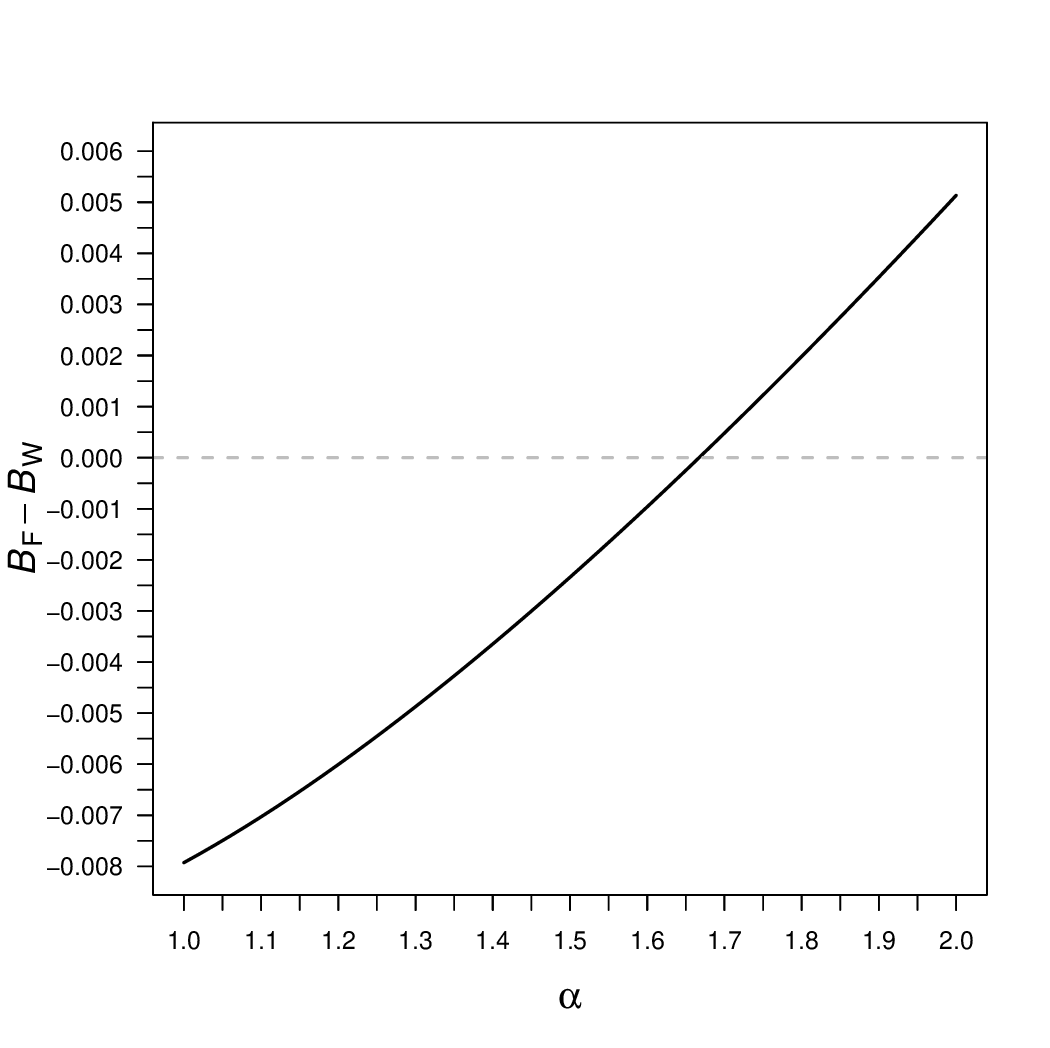,width=3in}
\label{ex3b}}
\caption{Aggregate investment for WD and FD as a function of $\alpha$ (left), and the difference, $B_{\rm F}-B_{\rm W}$, as a function of $\alpha$ (right).}
\label{fig:example3}
\end{center}
\end{figure}

\begin{example}
Suppose $n = 2$, $m = 3$, $\omega = 0.4$, and $F(v) = v^{\alpha}$. In Figure \ref{fig:example3}, we plot aggregate investment under FD (dashed lines) and under WD (solid lines) for values of $\alpha \in [1,2]$. Note that for $m = 3$, the elasticity of the distribution of types satisfies $\alpha > 1/(m-1)$, so that the condition in Proposition \ref{prop_comp_nd_fd} is satisfied. Figure \ref{fig:example3} shows that $B_{\rm W}$ may nevertheless be higher or lower than $B_{\rm F}$.
\end{example}

Yet, the above discussion implies that an unambiguous comparison between WD and FD can be made in the important special case of full entry (or, in other words, when the number of players in the contest is fixed). Since there is, generically, more entry under FD, a necessary and sufficient condition for full entry under all three disclosure rules is that $v^*=\underline{v}$, i.e., $\omega\leqslant\frac{(m-1)}{nm-1}\underline{v}$. Recall also that, from Proposition \ref{prop_comp_nd_wd}, WD dominates ND for any cutoff, including the full entry case. We, therefore, have the following result.

\begin{corollary}
\label{cor_fixed_n}
Suppose $m>1$ and there is full entry under all three disclosure rules (i.e., $\omega\leqslant\frac{(m-1)}{nm-1}\underline{v}$ or the number of players in the contest is fixed). Then expected aggregate investment under within-group disclosure is greater than under both no disclosure and full disclosure.
\end{corollary}

\section{Extensions}
\label{sec:extensions}

We consider four extensions. In Section \ref{sec:hetero_sizes}, we discuss the effect of disclosure for two groups of different sizes. Additive aggregation technology is discussed in Section \ref{sec:additive}. In Section \ref{sec:other_objectives}, we study the impact of disclosure policies on two alternative objectives of the principal---expected total investment and expected highest investment. In Section \ref{sec:ind_vs_group}, we compare group contests to individual contests among the same players under different disclosure rules and discuss when the principal would prefer either type.

\subsection{Two groups with different sizes}
\label{sec:hetero_sizes}

We consider $n=2$ groups of different sizes, $m_1$ and $m_2$. To simplify matters, we restrict attention to the case of full participation and focus on the impact of disclosure. Assume, without loss, that group 2 is larger, i.e., $m_1<m_2$.

\paragraph{No disclosure} The ND setting with two asymmetric groups is analyzed in Section 4 of \cite{Barbieri-Malueg:2016} (henceforth, BM16), which relies on the results of \cite{Amann-Leininger:1996}. We consider a semi-symmetric equilibrium in which all players in group $i=1,2$ bid according to an increasing function $b_i(v)$.\footnote{As observed by BM16, these bidding functions can be zero up to a threshold value of $v$ and are strictly increasing afterwards. We treat the equations below as holding in the strictly increasing region.} Let $\beta_i(t)$ denote the inverse of $b_i$. Following BM16, the inverse bidding functions satisfy the system of equations
\begin{align}
& \label{bid_sys_1} m_2\beta_1(t)F(\beta_1(t))^{m_1-1}F(\beta_2(t))^{m_2-1}f(\beta_2(t))\beta_2'(t) = 1, \\
& \label{bid_sys_2} m_1\beta_2(t)F(\beta_1(t))^{m_1-1}F(\beta_2(t))^{m_2-1}f(\beta_1(t))\beta_1'(t) = 1.\nonumber
\end{align}
It is also useful to introduce a ``matching'' function $\gamma(v) = \beta_2(b_1(v))$, which satisfies $\gamma'(v) = \beta_2'(b_1(v))b_1'(v)$ and $\gamma(\overline{v})=\overline{v}$. The system of equations above then gives, after some transformations (see BM16 proof of Proposition 3),
\[
\gamma'(v) = \frac{m_1\gamma(v)f(v)}{m_2vf(\gamma(v))},
\]
which is a first-order ordinary differential equation (ODE) subject to the initial condition $\gamma(\overline{v})=\overline{v}$. Let $\gamma_{\rm N}(v)$ denote the unique solution. The bidding functions can then be found by setting $\beta_1(t)=v$ in (\ref{bid_sys_1}), which turns it into 
\[
b_1'(v) = m_2vF(v)^{m_1-1}F(\gamma_{\rm N}(v))^{m_2-1}f(\gamma_{\rm N}(v))\gamma_{\rm N}'(v) = m_1\gamma_{\rm N}(v)f(v)F(v)^{m_1-1}F(\gamma_{\rm N}(v))^{m_2-1}.
\]
The bidding function in group 1 is, therefore,
\[
b_{1,{\rm N}}(v) = m_1\int_{\underline{v}}^v \gamma_{\rm N}(s)F(s)^{m_1-1}F(\gamma_{\rm N}(s))^{m_2-1}\dd F(s).
\]
Because $m_1<m_2$, Proposition 3 of BM16 implies that $b_{1,{\rm N}}(v)\geqslant b_{2,{\rm N}}(v)$ for all $v\in(\underline{v},\overline{v})$. Therefore, bidding in group 1 starts at the lower bound of the support of $F$; whereas bidding in group 2 starts at $\gamma_{\rm N}(\underline{v})\geqslant\underline{v}$. From symmetry, the bidding function in group 2 is
\[
b_{2,{\rm N}}(v) = m_2\int_{\gamma_{\rm N}(\underline{v})}^v \gamma_{\rm N}^{-1}(s)F(s)^{m_2-1}F(\gamma_{\rm N}^{-1}(s))^{m_1-1}\dd F(s), \quad v\geqslant \gamma_{\rm N}(\underline{v}).
\]
For $m_1=m_2$ and $\gamma_{\rm N}(v)=v$, these equations agree with the symmetric case. 

Expected aggregate investment in the contest then can be found as
$B_{\rm N} = B_{1,{\rm N}} + B_{2,{\rm N}}$, where $B_{i,{\rm N}} = \int b_{i,{\rm N}}(v)\dd F(v)^{m_i}$. These integrals can be simplified as
\begin{align*}
& B_{1,{\rm N}} = \int_{\underline{v}}^{\overline{v}}\int_{\underline{v}}^v m_1\gamma_{\rm N}(s)F(s)^{m_1-1}F(\gamma_{\rm N}(s))^{m_2-1}\dd F(s)\dd F(v)^{m_1} \\
& = \int_{\underline{v}}^{\overline{v}}\gamma_{\rm N}(s)F(\gamma_{\rm N}(s))^{m_2-1}[1-F(s)^{m_1}]\dd F(s)^{m_1}
\end{align*}
and
\begin{align*}
& B_{2,{\rm N}} = \int_{\gamma_{\rm N}(\underline{v})}^{\overline{v}}\gamma_{\rm N}^{-1}(s)F(\gamma_{\rm N}^{-1}(s))^{m_1-1}[1-F(s)^{m_2}]\dd F(s)^{m_2} \\
& = \int_{\underline{v}}^{\overline{v}}sF(s)^{m_1-1}[1-F(\gamma_{\rm N}(s))^{m_2}]\dd F(\gamma_{\rm N}(s))^{m_2}.
\end{align*}

\paragraph{Within-group disclosure} Under WD, the game reduces to a contest between two players with valuations distributed according to $F(v)^{m_i}$, $i=1,2$. We can use Eqns. (7) and (8) from BM16, setting $c_1=c_2=n_1=n_2=1$ and $F_i=F^{m_i}$, $i=1,2$:
\begin{align*}
& m_2\beta_1(t)F(\beta_2(t))^{m_2-1}f(\beta_2(t))\beta_2'(t) = 1,\\
& m_1\beta_2(t)F(\beta_1(t))^{m_1-1}f(\beta_1(t))\beta_1'(t) = 1.
\end{align*}
Furthermore, the matching function $\gamma(v)=\beta_2(b_1(v))$ now satisfies (17) from BM16,
\[
\gamma'(v) = \frac{m_1\gamma(v)F(v)^{m_1-1}f(v)}{m_2vF(\gamma(v))^{m_2-1}f(\gamma(v))},
\]
subject to the initial condition $\gamma(\overline{v})=\overline{v}$. We let $\gamma_{\rm W}(v)$ denote the unique solution. In this case, since $m_1<m_2$, the active bidder from group 2 has value distribution $F^{m_2}$ which first-order stochastically dominates $F^{m_1}$. Hence, the group 2 bidder is active starting from $\underline{v}$, whereas group 1 bidder is active from $\gamma_{\rm W}^{-1}(\underline{v})$.
Setting $\beta_1(t)=v$ in the first ODE above gives
\[
m_2vF(\gamma(v))^{m_2-1}f(\gamma(v))\frac{\gamma'(v)}{b_1'(v)} = 1,
\]
which produces
\[
b_1'(v) = m_1\gamma(v)F(v)^{m_1-1}f(v),
\]
and hence the bidding function in group 1 is
\[
b_{1,{\rm W}}(v) = m_1\int_{\gamma_{\rm W}^{-1}(\underline{v})}^v \gamma_{\rm W}(s)F(s)^{m_1-1}\dd F(s), \quad v\geqslant \gamma_{\rm W}^{-1}(\underline{v}).
\]
Symmetrically, for group 2 we have
\[
b_{2,{\rm W}}(v) = m_2\int_{\underline{v}}^v \gamma_{\rm W}^{-1}(s)F(s)^{m_2-1}\dd F(s). 
\]

As in the ND case, we can find expected aggregate investment as
$B_W = B_{2,{\rm W}} + B_{2,{\rm W}}$, where $B_{i,{\rm W}} = \int b_{i,{\rm W}}(v)\dd F(v)^{m_i}$. The two components are
\begin{align*}
& B_{1,{\rm W}} = \int_{\gamma_{\rm W}^{-1}(\underline{v})}^{\overline{v}}\int_{\gamma_{\rm W}^{-1}(\underline{v})}^v \gamma_{\rm W}(s)\dd F(s)^{m_1}\dd F(v)^{m_1} \\
& = \int_{\gamma_{\rm W}^{-1}(\underline{v})}^{\overline{v}}\gamma_{\rm W}(s)[1-F(s)^{m_1}]\dd F(s)^{m_1} = \int_{\underline{v}}^{\overline{v}}s[1-F(\gamma_{\rm W}^{-1}(s))^{m_1}]\dd F(\gamma_{\rm W}^{-1}(s))^{m_1}
\end{align*}
and, symmetrically,
\begin{align*}
B_{2,{\rm W}} = \int_{\underline{v}}^{\overline{v}}\gamma_{\rm W}^{-1}(s)[1-F(s)^{m_2}]\dd F(s)^{m_2}.
\end{align*}

\paragraph{Full disclosure} Finally, under FD we again have top valuation bidders from each group competing but this time their valuations, drawn from $F^{m_1}$ and $F^{m_2}$, are common knowledge. Therefore, expected aggregate investment is 
\[
B_F = \int_{t_2<t_1}\left(\frac{t_2}{2}+\frac{t_2^2}{2t_1}\right)\dd F(t_1)^{m_1} \dd F(t_2)^{m_2} + \int_{t_2<t_1}\left(\frac{t_2}{2}+\frac{t_2^2}{2t_1}\right)\dd F(t_1)^{m_2} \dd F(t_2)^{m_1}.
\]
Here, the first term covers cases where the top valuation in group 1 is larger than in group 2 and the second term covers the opposite.

In Appendix \ref{sec:appendix_hetero}, we provide explicit expressions for functions $\gamma_{\rm N}$ and $\gamma_{\rm W}$ as well as aggregate investment under ND, WD, and FD for $F(v)=v^{\alpha}$ on $[0,1]$. Our numerical experiments with this distribution and a range of parameters $(m_1,m_2,\alpha)$ show that $B_{\rm W}>B_{\rm N}$ holds throughout; at least, we were not able to find a counterexample. Proving this in general, however, appears difficult even in this special case. First, it can be shown that aggregate investment in group 1 is lower under WD, $B_{1,{\rm W}}<B_{1,{\rm N}}$, whereas aggregate investment in group 2 is higher under WD, $B_{2,{\rm W}}>B_{2,{\rm N}}$. Hence, the overall comparison requires showing that $B_{2,{\rm W}}-B_{2,{\rm N}}>B_{1,{\rm N}}-B_{1,{\rm W}}$, i.e., that the increase for the larger group dominates the decrease for the smaller one. Second, the matching functions $\gamma_{\rm N}$ and $\gamma_{\rm W}$ are not ordered pointwise, even for $F(v)=v^{\alpha}$. Yet, given our numerical results, it is reasonable to conjecture that $B_{\rm W}>B_{\rm N}$ holds in general---a question we leave for future research.\footnote{We also find $B_{\rm W}>B_{\rm N}$ for truncated Pareto distributions similar to those considered in Example \ref{ex_Bmax}.}

We performed similar computations comparing FD to ND and FD to WD. We find the results of these comparisons to be very similar to the homogeneous case: The impact of full disclosure is likewise ambiguous and can be reversed depending on the elasticity of the distributions of values. Similar to Proposition \ref{prop_comp_nd_fd}, $B_{\rm F}>B_{\rm N}$ when $F$ is sufficiently elastic, and the opposite holds when $F$ are sufficiently inelastic. We also observe that $B_{\rm W}>B_{\rm F}$, which is consistent with Corollary \ref{cor_fixed_n} under full entry.

We conclude that, perhaps surprisingly, our results under the homogeneity assumption are somewhat robust; at least in the two-group heterogeneous case, we were not able to find qualitatively different results in numerical experiments for different group sizes and value distributions.

This is, to a certain extent, intuitive. The effects of full disclosure we identified in the homogeneous case are still present in the heterogeneous case. Likewise, within-group disclosure solves the within-group coordination problem regardless of whether groups are heterogeneous, which means that group leaders no longer need to shade their bids in anticipation of the possibility that they are not the highest valuation bidders. However, our numerical results also show that in the smaller group this leads to a reduction in investment. Thus, strategic interaction between groups breaks this simple intuition, and the overall effect is nontrivial.

\subsection{Additive aggregation technology}
\label{sec:additive}

We consider a model in which within-group aggregation is additive. For simplicity, we ignore entry and assume that all players participate. That is, group $i$ wins if $\sum_{j=1}^mb_{ij}>\sum_{j=1}^mb_{kj}$ for all $k\ne i$, with zero-probability ties broken in an arbitrary manner. 

In this setting, it is well known that equilibrium characterization in the ND environment is prohibitively difficult \citep[for a discussion see, for example,][henceforth, BT25]{Barbieri-Topolyan:2025}. The main difficulty is in computing the distribution of aggregate group bids that is a multi-dimensional convolution including the unknown bidding function. BT25 circumvent this problem by considering correlated group-level equilibria where players within each group condition their bids on a common signal from a randomization device. Notably, BT25 assume a \emph{strictly concave} CES aggregation rule excluding perfect substitutes; that is, even their correlated equilibrium approach is not suitable for the additive aggregation setting.

At the same time, the WD and FD environments are essentially equivalent to the best-shot aggregation case. Indeed, in both cases the highest valuation bidder will be the only active player in each group with probability one, and competition between groups then has the same equilibrium characterization as before. Thus, the WD and FD environments solve the coordination problem exactly as before, while the ND environment is subject to free riding. Of course, it is not clear \emph{a priori} how additive bidding and free riding will work together, which makes the comparison between ND and WD nontrivial.

To fix matters, it is reasonable to look for a symmetric equilibrium in the form of a monotone bidding function $b(v)$. For a player with valuation $v$ pretending to be $\hat{v}$, the expected payoff is then given by
\[
\pi(v,\hat{v}) = v\mathds{P}\left[b(\hat{v})+\sum_{j>1}b(v_{1j})>\sum_{j\geqslant 1}b(v_{kj}) \text{ for all }k> 1\right] - b(\hat{v}),
\]
where the probability is calculated with respect to the distribution of $v_{1j}$, $j=2,\ldots,m$, and $v_{kj}$, $k=2,\ldots,n$, $j=1,\ldots,m$. This payoff function can be written as
\[
\pi(v,\hat{v}) = vG[\hat{v};b(\cdot)] - b(\hat{v}),
\]
where $G[\cdot;b(\cdot)]$ is the distribution of $b^{-1}(\max_{k=2,\ldots,n}\sum_{j\geqslant 1}b(v_{kj})-\sum_{j>1}b(v_{1j}))$. 

Distribution $G$ can in principle be written as a multi-dimensional convolution integral that includes function $b(\cdot)$ together with its inverse and derivative. The equilibrium first-order condition then takes the form $b'(v) = vg[v;b(\cdot)]$, where $g=G'$ is the associated density. The first-order condition, therefore, is a complex integro-differential equation for $b(\cdot)$. There are no known methods to solve it even in the simplest case, such as $n=m=2$ and $F$ uniform. Moreover, it has not been established whether $\pi(v,\hat{v})$ satisfies the usual single-crossing property, i.e., whether the first-order condition provides a global maximum.

Instead of solving the first-order condition, we  have created, with help from ChatGPT, a Python program that directly computes the symmetric equilibrium bidding function $b(\cdot)$ in the $n=m=2$ case using an iterative procedure. The algorithm starts by positing a candidate function $b_0(\cdot)$ on a grid of $v$ values, then empirically computes $G[\cdot;b_0(\cdot)]$ by simulating the values of $n-1$ other bidders, and calculates for each $v$ in the grid the best response bid $b_1(v):=b_{\rm br}[v;b_0(\cdot)]$. This best response is then used to compute the next iteration of $G$, and the procedure is repeated  until convergence is reached. Our numerical experiments with various distributions of values showed total expected effort under WD above ND in all cases. Of course, given the limited nature of the experiments it is difficult to make general claims, but the dominance of WD over ND so far looks as a reasonable conjecture in this setting.

\subsection{Alternative objectives}
\label{sec:other_objectives}

\subsubsection{Total investment}

The ``best-shot'' production technology within groups implies that the investments of group members that are below the maximum bid are essentially wasted. However, from the contest designer's or policy perspective it may be of interest to consider \textit{total investment}, defined as the sum of all individual investments, as a relevant criterion. While the wasted portion of total investment does not contribute directly to output (e.g., it does not improve the quality of the resulting innovation produced by the group), it may have spillovers the designer cares about. From (\ref{b_nd}), the expected total investment by entrants under no disclosure is
\begin{equation}
\label{B_nd_tot}
B^{\rm tot}_{\rm N} = nmq\int_{v^*}^{\overline{v}}b_{\rm N}(v)\dd\tilde{F}(v) = m^2n(n-1)\int_{t_1\geqslant t_2\geqslant v^*}t_2F(t_2)^{nm-2}\dd F(t_1)\dd F(t_2).
\end{equation} 

Notice that under both within-group disclosure and full disclosure, expected total investment coincides with expected aggregate investment, Eqs. (\ref{B_wd}) and (\ref{B_fd}), respectively, because only leaders are (potentially) active in each group. Comparing (\ref{B_nd_tot}) and (\ref{B_wd}), we obtain 
\begin{align*}
&B_{\rm W}-B^{\rm tot}_{\rm N}\\
& = m^2n(n-1)\int_{t_1\geqslant t_2\geqslant v^*}t_2F(t_2)^{nm-m-1}F(t_1)^{m-1}\left[F(t_1)^{m-1}-F(t_2)^{m-1}\right]\dd F(t_1)\dd F(t_2)\geqslant 0.
\end{align*}
Thus, aggregate (equivalently, total) investment under within-group disclosure exceeds total investment without disclosure. The inequality is strict for $m>1$.\footnote{Recall that $m=1$ implies WD and ND collapse to the same information environment.} With disclosure, the entrants are able to solve the coordination problem within groups and bid more effectively. The impact of within-group disclosure is so strong that the resulting expected aggregate investment exceeds even the sum of individual investments under no disclosure.

This result also implies that the comparison of total investments between ND and FD is ambiguous. Specifically, we have $B_{\rm W}>B^{\rm tot}_{\rm N}>B_{\rm N}$, while both $B_{\rm W}$ and $B_{\rm N}$ can be either above or below $B_{\rm F}$.\footnote{The ambiguity of comparison between $B_{\rm F}$ and $B^{\rm tot}_{\rm N}$ can also be seen directly from the proof of Proposition \ref{prop_comp_nd_fd} in Appendix \ref{sec:appendix_Eq15}, by observing that the first inequality contains the difference between them. Therefore, the sufficient condition of Proposition \ref{prop_comp_nd_fd} applies to this comparison as well.}

\subsubsection{Expected highest investment}

In an innovation or R\&D competition setting, the principal may be interested in maximizing the expected highest investment across all participating groups. For example, if each group submits a solution to a challenge such as the Netflix Prize\footnote{See \url{https://www.thrillist.com/entertainment/nation/the-netflix-prize}.} or the XPRIZE Carbon Removal initiative,\footnote{See \url{https://www.xprize.org/prizes/carbonremoval}.} only the best solution (if any) will eventually be implemented on a large scale. In this section we compare \emph{expected highest investment} across the disclosure policies.

Under no disclosure, Eq. (\ref{b_nd}) gives the expected highest investment 
\begin{align}
\label{B_nd_max}
& B^{\max}_{\rm N} = \sum_{k=1}^{nm}\binom{nm}{k}q^k(1-q)^{nm-k}\int_{v^*}^{\overline{v}}b_{\rm N}(v)\dd\tilde{F}(v)^k = \int_{v^*}^{\overline{v}}b_{\rm N}(v)\dd F(v)^{nm} \nonumber\\
& = m^2n(n-1)\int_{t_1\geqslant t_2\geqslant v^*}t_2F(t_2)^{nm-2}F(t_1)^{nm-1}\dd F(t_1)\dd F(t_2).
\end{align}

Under full disclosure, suppose there are at least two active groups, and $v_{(1)}>v_{(2)}$ are the two highest valuations. In equilibrium, the two entrants will bid according to the mixed strategies $G_1$ and $G_2$ described in Section \ref{sec:individuals}. Expected highest investment conditional on $v_{(1)}$ and $v_{(2)}$ then is
\begin{align*}
& \int_{x_1>x_2}x_1\dd G_1(x_1)\dd G_2(x_2) +  \int_{x_1<x_2}x_2\dd G_1(x_1)\dd G_2(x_2) \\
& = \int x_1G_2(x_1)\dd G_1(x_1) + \int x_2G_1(x_2)\dd G_2(x_2) \\
& = \int_0^{v_{(2)}}x_1\left(1-\frac{v_{(2)}}{v_{(1)}}+\frac{x_1}{v_{(1)}}\right)\frac{\dd x_1}{v_{(2)}} + \int_0^{v_{(2)}}\frac{x_2^2}{v_{(2)}}\frac{\dd x_2}{v_{(1)}} = \frac{v_{(2)}}{2} + \frac{v_{(2)}^2}{6v_{(1)}}.
\end{align*}
Using the joint distribution of $(v_{(1)},v_{(2)})$ from (\ref{B_fd}), we obtain the unconditional expected highest investment:
\begin{align}
\label{B_fd_max}
B^{\max}_{\rm F} = m^2n(n-1)\int_{t_1\geqslant t_2\geqslant v^*_{\rm F}}\left(\frac{t_2}{2}+\frac{t_2^2}{6t_1}\right)F(t_2)^{nm-m-1}F(t_1)^{m-1}\dd F(t_1)\dd F(t_2).
\end{align}

The comparison between (\ref{B_nd_max}) and (\ref{B_fd_max}) is ambiguous in general, but we establish the following result (for a proof, see Appendix \ref{sec:appendix_proofProp5}).

\begin{proposition}
\label{prop_comp_nd_fd_max}
Suppose there is full entry (i.e., $v^*=v^*_{\rm F}=\underline{v}$) and $\overline{v}$ is finite. Then $B^{\max}_{\rm F}>B^{\max}_{\rm N}$ for $n$ or $m$ sufficiently large.
\end{proposition}
The reason why full disclosure dominates for sufficiently large $n$ or $m$ is that, as $n$ or $m$ increases, bids are shaded more under no disclosure because the probability for a player's bid to be the winning bid for a given prize valuation declines. This effect is absent under full disclosure where the two highest valuation players know that they are the only potential bidders. As the number of players increases, bid shading eventually undermines the expected highest investment in ND and, to a smaller extent, in WD, as compared to FD. Of course, under partial entry there is additional ambiguity due to the difference between the entry cutoffs. The assumption of bounded support is critical for the above argument, however, because the top order statistics converge to $\overline{v}$ at different speeds under ND and FD. As seen from (\ref{B_nd_max}) and (\ref{B_fd_max}), the speed of convergence is generally higher under ND. This implies that if the support of $F$ is not bounded and the distribution has a sufficiently heavy tail, $B^{\max}_{\rm N}$ can dominate $B^{\max}_{\rm F}$.

\begin{figure}[H]
\begin{center}
\subfloat{\epsfig{figure=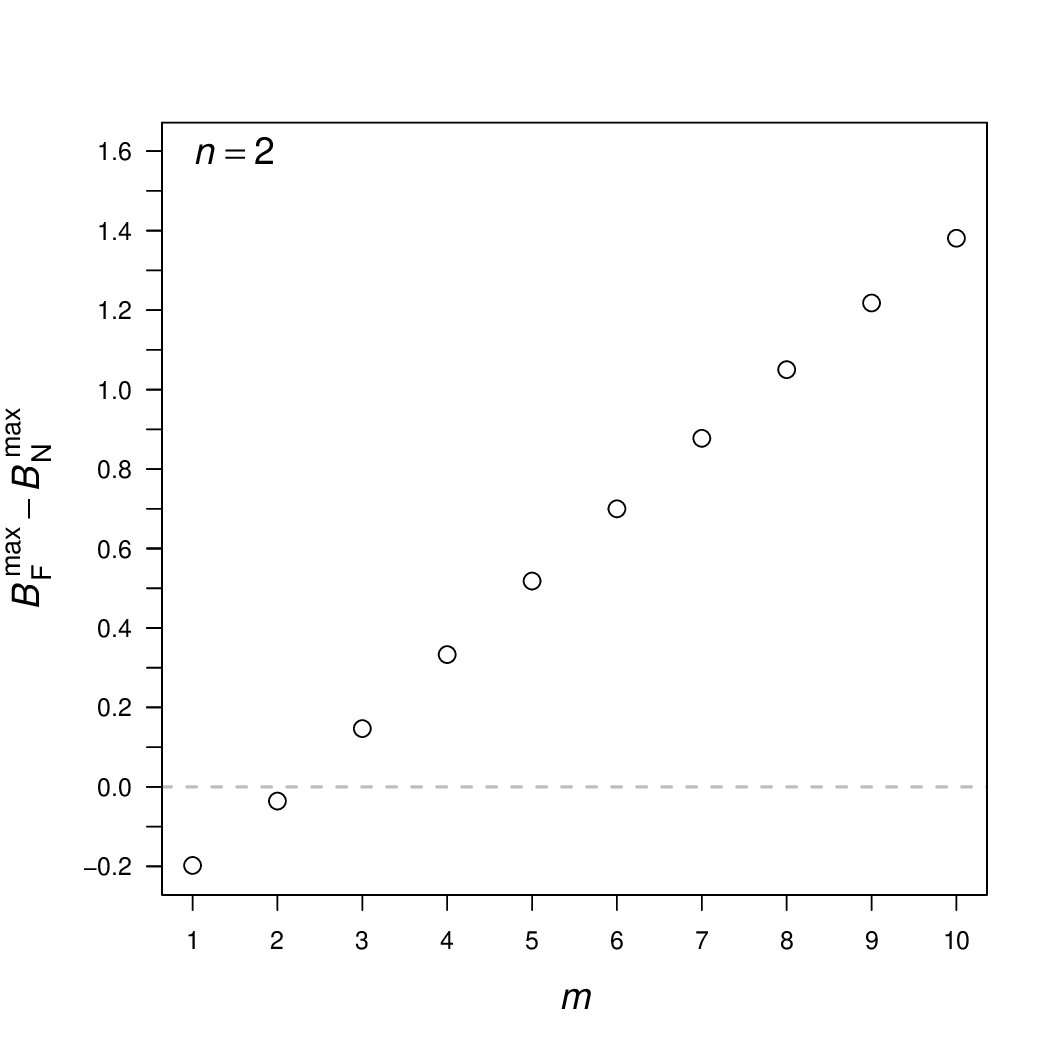,width=3in}}
\subfloat{\epsfig{figure=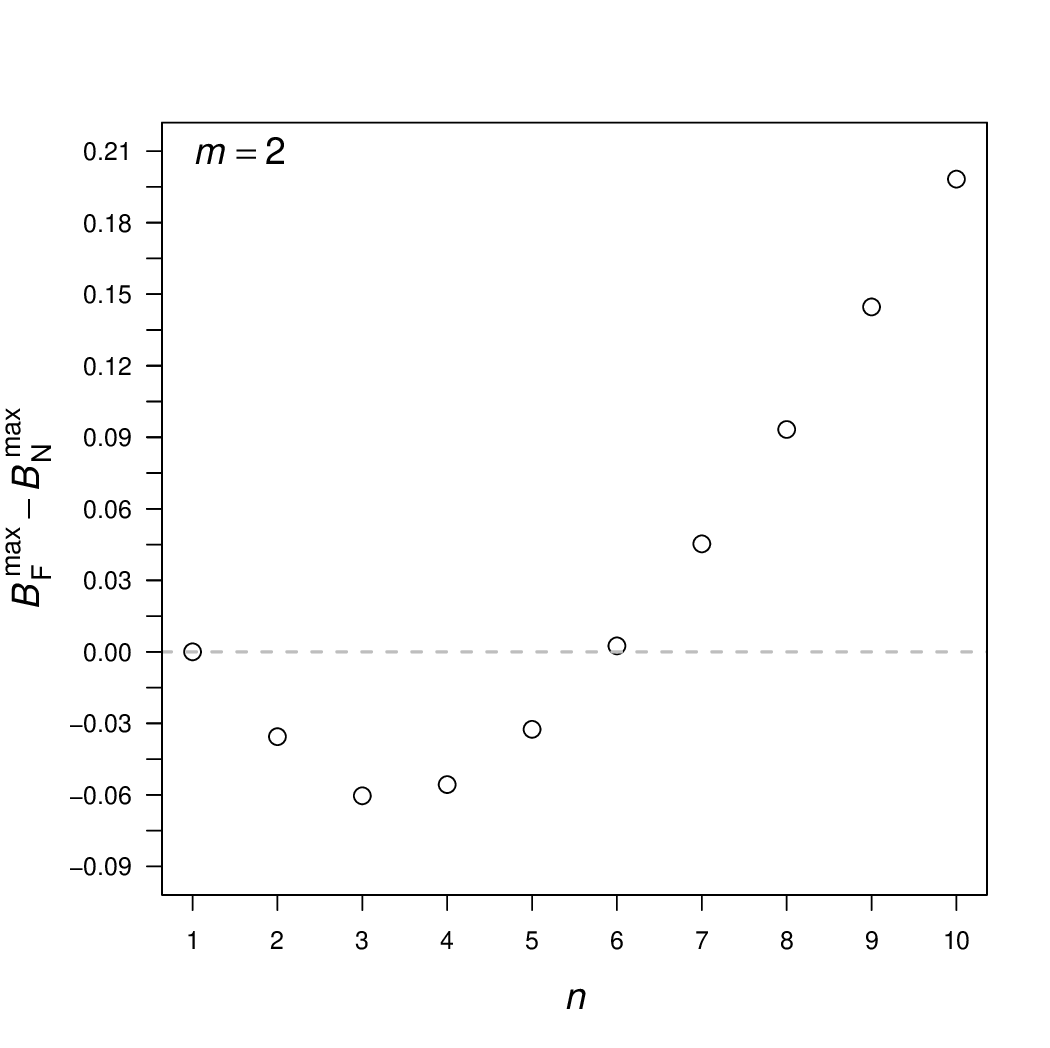,width=3in}}
\vskip -1.0cm
\subfloat{\epsfig{figure=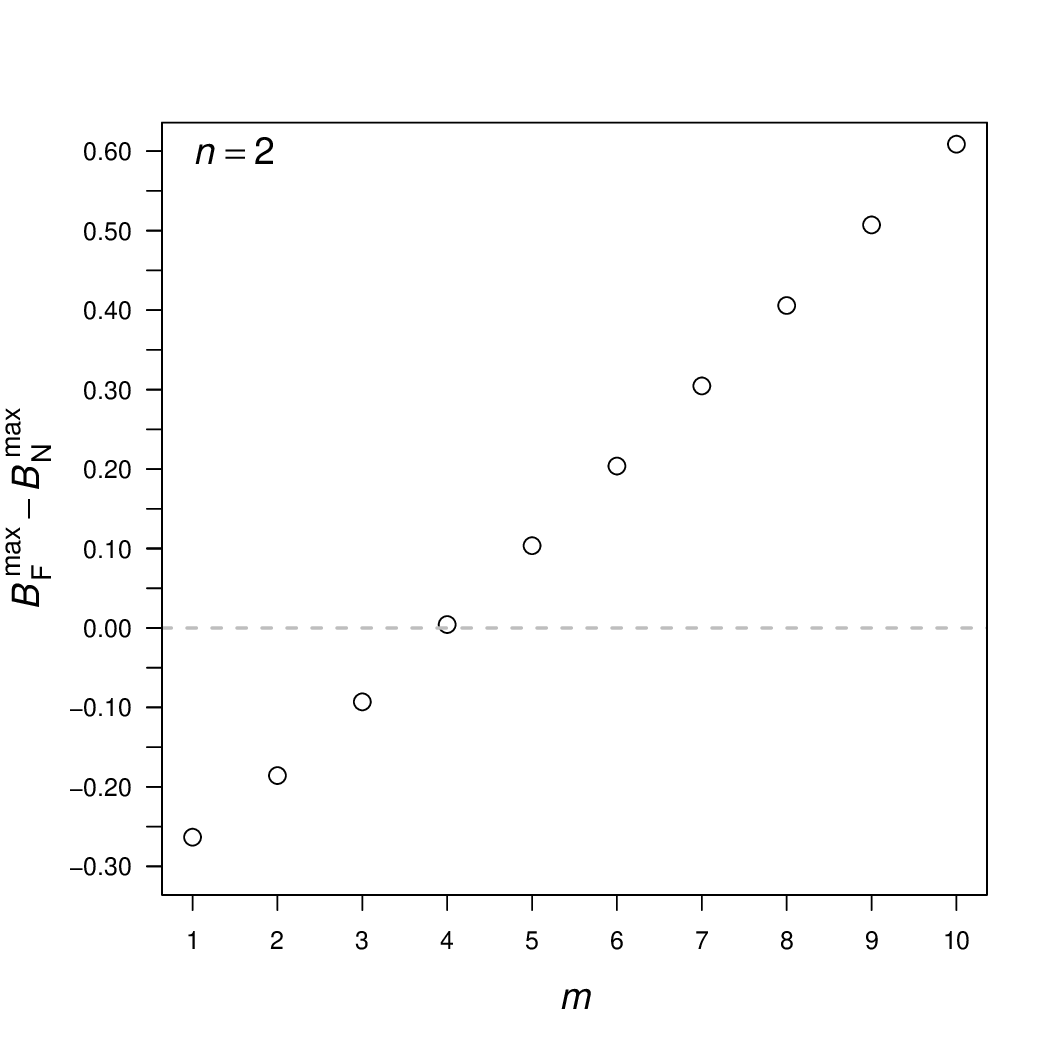,width=3in}}
\subfloat{\epsfig{figure=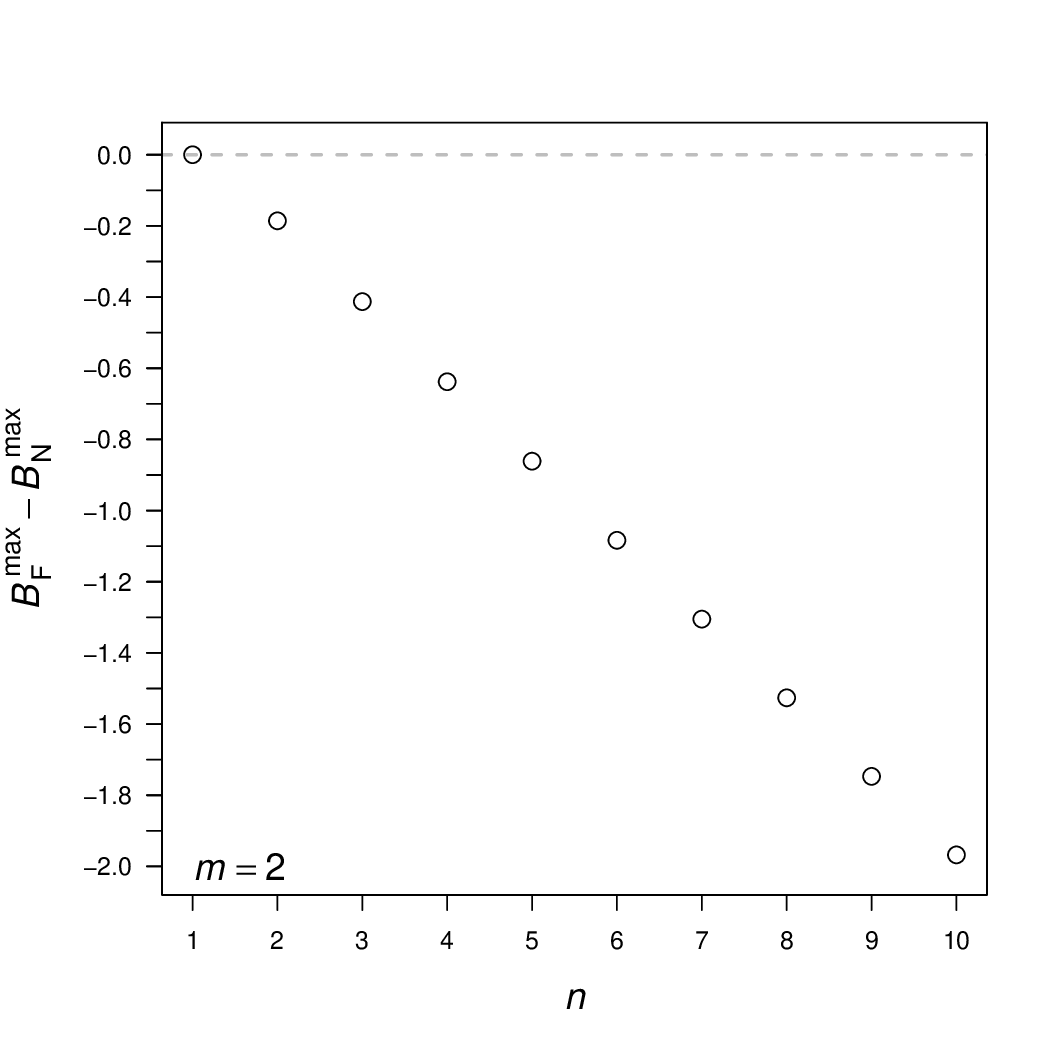,width=3in}}
\caption{The difference in expected highest investment between FD and ND, $B^{\max}_{\rm F}-B^{\max}_{\rm N}$, as a function of $m$ for $n=2$ (the left panels), and as a function of $n$ for for $m=2$ (the right panels). The distribution of values is truncated Pareto (the top two panels) and Pareto (the bottom two panels), with $p=1$ and $v_{\max}=30$.}
\label{fig:max_example}
\end{center}
\end{figure}

\begin{example}
\label{ex_Bmax}
Suppose $\omega=0$, i.e., there is full entry under ND and FD. To illustrate Proposition \ref{prop_comp_nd_fd_max}, we consider Pareto distribution, $F(v)=1-\frac{1}{(v+1)^p}$, with support $[0,\infty)$, and its truncated version, $F_{\rm t}(v)=\frac{F(v)}{F(v_{\max})}$, supported in $[0,v_{\max}]$. Figure \ref{fig:max_example} shows the dependence of the difference $B^{\max}_{\rm F}-B^{\max}_{\rm N}$ on $m$, for $n$ fixed, and on $n$, for $m$ fixed. The support of $F_{\rm t}$ is bounded, and $B^{\max}_{\rm F}$ eventually dominates $B^{\max}_{\rm N}$ for $m$ or $n$ large enough. In contrast, $B^{\max}_{\rm N}$ dominates $B^{\max}_{\rm F}$ for $n$ large (and $m$ fixed) when the support is not bounded. Note that $B^{\max}_{\rm F}$ still dominates $B^{\max}_{\rm N}$ for $m$ large (and $n$ fixed), illustrating that the bounded support condition is sufficient but not necessary.
\end{example}

A clear comparison of expected highest investment can be made between ND and WD. Indeed, as seen from Eqs. (\ref{b_nd}) and (\ref{b_wd}), the equilibrium bidding functions $b_{\rm N}(v)$ and $b_{\rm W}(v)$ are ranked pointwise such that $b_{\rm W}(v)\geqslant b_{\rm N}(v)$ for all $v\in V^*$, with strict inequality for $m>1$ and $v>v^*$. Since the distribution of the highest valuation is the same in both cases, the highest investment in WD exceeds the highest investment in ND in the sense of first-order stochastic dominance. It is, therefore, immediate that expected highest investment is higher under WD. The effect becomes stronger as $m$ increases because players benefit more from solving the coordination problem with increasing group size.

\subsection{Group contests vs. individual contests}
\label{sec:ind_vs_group}

From a contest design perspective, it might be of interest to compare group contests to individual contests for the same set of players.\footnote{We thank an anonymous reviewer for suggesting this extension.} To this end, suppose there are $N=nm$ players, and the contest designer can either run an individual contest or a group contest splitting the players into $n$ groups. For simplicity, we assume throughout this section that there is full entry in all cases (for example, because $\omega=0$), i.e., $v^*=v^*_{\rm F}=\underline{v}$. Under ND, aggregate investment in the individual contest is then given by (\ref{B_nd_ind}) where $n$ is replaced with $nm$:
\begin{align}
\label{B_indND_comp}
B_{\rm I,N} = nm(nm-1)\int_{t_1\geqslant t_2}t_2F(t_2)^{nm-2}\dd F(t_1)\dd F(t_2).
\end{align}
Comparing this expression to (\ref{B_nd}) for the group contest, we observe that the integrand in the former is always greater because $F(t_1)^{m-1}\leqslant 1$; and the ratio of the coefficients in front of the integrals is $\frac{nm-1}{m(n-1)}\geqslant 1$. We conclude that individual contests produce a higher aggregate investment. This is rather intuitive, given that players in a group can free ride on their group's leader, and there is potential for their bids to be wasted. More generally, this result also suggests that splitting individuals into smaller groups would raise aggregate investment. 

The situation changes under WD where aggregate investment in the individual contest is still given by (\ref{B_indND_comp}), but in the group contest it is (\ref{B_wd}). The integrand in the latter is now \emph{larger} because 
\[
F(t_2)^{nm-m-1}F(t_1)^{m-1}\geqslant F(t_2)^{nm-m-1}F(t_2)^{m-1} = F(t_2)^{nm-2}
\]
in the domain of integration. The coefficient in front of the integral is still larger in (\ref{B_indND_comp}), but the ratio of the coefficients tends to 1 as $n$ increases. Thus, at least for $N$ large enough, when many groups can be created, the group contest among many groups outperforms the individual one in terms of aggregate investment. The effect is due to coordination within groups which allows group leaders to effectively participate in an individual contest among themselves. The valuations of these group leaders first-order stochastically dominate those in the general population (i.e., in the individual contest), and the impact of this dominance overcomes the pure numbers advantage of the individual contest, at least when there are sufficiently many groups.

Finally, under FD, aggregate investment in the individual contest is given by (\ref{B_fd_ind}) where $n$ is replaced by $nm$:
\[
B_{\rm I,F} = nm(nm-1)\int_{t_1\geqslant t_2}\left(\frac{t_2}{2}+\frac{t_2^2}{2t_1}\right)F(t_2)^{nm-2}\dd F(t_1)\dd F(t_2).
\]
Comparing this to (\ref{B_fd}) for the group contest, we observe the same forces at play: The integrand is larger in (\ref{B_fd}), and the coefficient in front of the integral is larger for the individual contest; and again the ratio of the coefficients tends to 1 for $n$ large. Thus, the group contest yields a larger aggregate investment when there are sufficiently many groups under full disclosure as well.

We conclude that disclosure has important effects on contest design, in an environment where the designer can choose whether to run an individual or a group contest or, more generally, can split a set of players into groups. Without disclosure, such grouping is not beneficial, but with disclosure, when sufficiently many groups can be created, having coordination within groups helps boost aggregate investment as compared to the individual contest.

\section{Concluding remarks}\label{sec:conc}
This paper is the first to study endogenous entry in contests between groups. Our focus is on the effects of information disclosure, which can be a designer's choice or institutionally predetermined. 
In settings where winning the contest provides a group-specific public good, group contests generate an interesting interplay of the standard contest incentives across groups with social dilemma-type free-riding incentives within groups. Endogenous entry and information disclosure can affect those incentives in several ways, leading to higher or lower aggregate equilibrium investment as compared to the no disclosure benchmark. The direction of the effect depends on how information is disclosed, the group size, and on the properties of the distribution of types. 

Without disclosure, the competition between groups is efficient, in that the prize is allocated to the group containing the player with the highest value. However, significant investment within groups is wasted due to lack of coordination. Within-group disclosure, under which players only find out about the number and types of other members of their own group, helps players within groups coordinate on more socially efficient investment strategies at the group level. Competition across groups remains efficient, and the marginal type does not change, producing a higher aggregate investment overall.

Full disclosure, under which the information about all contest participants is disclosed across groups, helps within-group coordination as well, but produces two additional (and possibly competing) effects. First, the payoff of the marginal type, and hence the mass of players entering the contest, increases. This is because under no disclosure (or under within-group disclosure) the marginal type can only win the contest with a positive probability if she is the only entrant or the highest value entrant in her group has the highest value overall. In contrast, under full disclosure the marginal type can also win with a positive probability when the highest value entrant in her group has the second highest value overall. This additional probability of winning is linked to the inefficiency of the mixed equilibrium in all-pay auctions under complete information.

Second, depending on the properties of the distribution of types, bidding conditional on type may increase or decrease under full disclosure as compared to no disclosure. We show that full disclosure produces a higher aggregate investment when the distribution of types is sufficiently elastic. The condition becomes weaker as the group size increases. Intuitively, when the density of types increases at the upper bound of its support, the top order statistics of types are sufficiently close, and hence the reduction in bidding due to the second highest type dropping out is less severe, more so the larger the (expected) group size.

One clear implication of our results is that under full entry within-group disclosure is beneficial and should be facilitated by the contest designer when possible. This is true not only in cases where the designer's goal is to maximize expected aggregate investment, but also for a designer whose goal is to maximize expected total investment or expected highest investment. At the same time, disclosure in the public domain, such as the various ``leaderboard'' practices in crowdsourcing, or sunshine laws in government practices, should be exercised with caution, more so the smaller the group size. However, in the presence of uncertainty about the number of entrants, full disclosure may be beneficial in some cases. 

In particular, while for individual contests full disclosure unambiguously leads to a lower investment and should be avoided,\footnote{For a fixed number of players, this goes back to \cite{Morath-Munster:2008}.} it can become optimal as the group size increases, especially if the distribution of types is sufficiently elastic. Our results, therefore, suggest that different disclosure policies can be optimal depending on whether a competitive task is performed by individuals, small groups or large groups, even if other features of the environment are similar.

To summarize, our results indicate that the dominance of one information structure over another is nuanced and depends on both the organizer's objective and environmental factors such as the underlying distribution of types. When the goal is to maximize expected aggregate investment, the designer should generally prefer an intermediate level of disclosure that reveals information about types within groups, but not across groups. Such a structure outperforms a setting with no information about others' types, although it does not necessarily dominate a setting with full information. A similar pattern emerges when considering alternative objectives, such as maximizing expected total investment.

Relative to the intermediate disclosure regime, the potential benefits of full disclosure can be offset by a discouragement effect: when contestants become fully informed about competitors' types, the second-most aggressive contestant---the only other active player---may substantially reduce their investment or effectively withdraw from competition. Consequently, full disclosure is most attractive when this discouragement effect is sufficiently weak, which is more likely in environments where the strongest competitors tend to be similar in type or ability.


\bibliographystyle{aea}

\appendix

\section{Missing derivations and proofs}
\label{sec:appendix}

\subsection{Equation (\protect\ref{b_nd})}
\label{sec:appendix_Eq6}

The equation $\Pi_{\hat{v}}(v,v;v^*)=0$ has the form $b'(v)=vp'(v)$, where, from (\ref{p_nd}),
\begin{align*}
&p'(v) = \sum_{k_1=0}^{m-1}\sum_{k_2=1}^{nm-m}\binom{m-1}{k_1}\binom{nm-m}{k_2}q^{k_1+k_2}(1-q)^{nm-1-k_1-k_2}\tilde{F}(v)^{k_1+k_2-1}\tilde{f}(v)k_2\\
& = q\tilde{f}(v)\sum_{k_1=0}^{m-1}\binom{m-1}{k_1}(q\tilde{F}(v))^{k_1}(1-q)^{m-1-k_1}
\sum_{k_2=1}^{nm-m}\binom{nm-m}{k_2}k_2(q\tilde{F}(v))^{k_2-1}(1-q)^{nm-m-k_2}\\
& \left.= f(v)[q\tilde{F}(v)+1-q]^{m-1}\frac{\partial(z+1-q)^{nm-m}}{\partial z}\right|_{z=q\tilde{F}(v)} = m(n-1)F(v)^{nm-2}f(v).
\end{align*}
This produces the equation $b'(v) = m(n-1)vF(v)^{nm-2}f(v)$, whose unique solution with initial condition $b(v^*)=0$ is given by (\ref{b_nd}).

\subsection{Equation (\protect\ref{p_nd_marginal})}
\label{sec:appendix_Eq7}

We write the sum in (\ref{p_nd_marginal}) as
\begin{align*}
& \int_0^q\sum_{k_1=0}^{m-1}\sum_{k_2=1}^{nm-m}\binom{m-1}{k_1}\binom{nm-m}{k_2}z^{k_1+k_2-1}(1-q)^{nm-1-k_1-k_2}k_1dz\\
& = \int_0^q \sum_{k_1=0}^{m-1}\binom{m-1}{k_1}k_1z^{k_1-1}(1-q)^{m-1-k_1}\sum_{k_2=1}^{nm-m}\binom{nm-m}{k_2}z^{k_2}(1-q)^{nm-m-k_2}dz\\
& = \int_0^q \frac{\partial (z+1-q)^{m-1}}{\partial z}\left[(z+1-q)^{nm-m}-(1-q)^{nm-m}\right]dz\\
& = (m-1)\int_0^q(z+1-q)^{nm-2}dz - (1-q)^{nm-m}[1-(1-q)^{m-1}]\\
& = \frac{m-1}{nm-1}[1-(1-q)^{nm-1}]-(1-q)^{nm-m}+(1-q)^{nm-1}\\
& = \frac{m-1}{nm-1} + \frac{m(n-1)}{nm-1}(1-q)^{nm-1}-(1-q)^{nm-m}.
\end{align*}
Combining with the first term in (\ref{p_nd_marginal}), obtain the result.

\subsection{Non-leaders in WD cannot profitably deviate from bidding zero}

\label{sec:appendix_Eq9}

Suppose bidder 1, with valuation $v_1$ and bidding the equilibrium $b_1$, is the highest valuation bidder in some group, and bidder 2 from the same group, with valuation $v_2<v_1$, is considering bidding $b_2>0$. Clearly, any $b_2\le b_1$ is strictly dominated by $b_2=0$, so assume $b_2>b_1$. Note that $b_1$ maximizes bidder 1's payoff, $v_1p(b_1)-b_1$, where $p(\cdot)$ is the probability of the group winning as a function of its (best-shot) bid; therefore, $v_1p(b_2)-b_2\le v_1p(b_1)-b_1$ and hence 
\begin{align*}
& v_2p(b_2)-b_2 \le v_2p(b_2) + v_1p(b_1)-v_1p(b_2)- b_1 \\
& = v_2p(b_1) + (v_1-v_2)p(b_1)-(v_1-v_2)p(b_2) - b_1 \\
& = v_2p(b_1) -(v_1-v_2)(p(b_2)-p(b_1)) - b_1 < v_2p(b_1),
\end{align*}
which shows that bidder 2 is better off with $b_2=0$.

\subsection{Equation (\protect\ref{p_fd_marginal})}

\label{sec:appendix_Eq12}

Start by performing integration over $t_3$ and summation over $k_3$ in (\ref{p_fd_m}). Integration over $t_3$ is on $[v^*,t_2]$ in the first integral and on $[v^*,t_1]$ in the second integral, producing
\begin{align*}
& p_{\rm F}(v^*) = (1-q)^{nm-m} + (n-1)\sum_{k_1=0}^{m-1}\sum_{k_2=1}^m\sum_{k_3=0}^{nm-2m}\binom{m-1}{k_1}\binom{m}{k_2}\binom{nm-2m}{k_3}\times\\
&\times q^{k_1+k_2+k_3}(1-q)^{nm-1-k_1-k_2-k_3}\left[\int_{t_1\geqslant t_2\geqslant v^*}\left(1-\frac{t_2}{2t_1}\right)\tilde{F}(t_2)^{k_3}\dd\tilde{F}(t_1)^{k_1}\dd\tilde{F}(t_2)^{k_2} + \right.\\
&\left. + \int_{t_2\geqslant t_1\geqslant v^*}\frac{t_1}{2t_2}\tilde{F}(t_1)^{k_3}\dd\tilde{F}(t_1)^{k_1}\dd\tilde{F}(t_2)^{k_2} \right] =\\
& = (1-q)^{nm-m} + (n-1)\sum_{k_1=0}^{m-1}\sum_{k_2=1}^m\binom{m-1}{k_1}\binom{m}{k_2}\times\\
&\times q^{k_1+k_2}(1-q)^{2m-1-k_1-k_2}\left[\int_{t_1\geqslant t_2\geqslant v^*}\left(1-\frac{t_2}{2t_1}\right)F(t_2)^{nm-2m}\dd\tilde{F}(t_1)^{k_1}\dd\tilde{F}(t_2)^{k_2} + \right.\\
&\left. + \int_{t_2\geqslant t_1\geqslant v^*}\frac{t_1}{2t_2}F(t_1)^{nm-2m}\dd\tilde{F}(t_1)^{k_1}\dd\tilde{F}(t_2)^{k_2} \right].
\end{align*}
Next, we sum over $k_1$ and $k_2$, and swap the variables of integration in the second integral:
\begin{align*}
& p_{\rm F}(v^*) = (1-q)^{nm-m} + (n-1)\left[\int_{t_1\geqslant t_2\geqslant v^*}\left(1-\frac{t_2}{2t_1}\right)F(t_2)^{nm-2m}\dd F(t_1)^{m-1}\dd F(t_2)^m + \right.\\
&\left. + \int_{t_1\geqslant t_2\geqslant v^*}\frac{t_2}{2t_1}F(t_2)^{nm-2m}\dd F(t_2)^{m-1}\dd F(t_1)^m \right].
\end{align*}

Separate the first integral into two parts and combine its second part with the second integral:
\begin{align}
\label{p_fd_rep1}
& p_{\rm F}(v^*) = (1-q)^{nm-m} + (n-1)\int_{t_1\geqslant t_2\geqslant v^*}F(t_2)^{nm-2m}\dd F(t_1)^{m-1}\dd F(t_2)^m + \\
&+ m(m-1)(n-1)\int_{t_1\geqslant t_2\geqslant v^*}\frac{t_2}{2t_1}F(t_1)^{m-2}F(t_2)^{nm-m-2}[F(t_1)-F(t_2)]\dd F(t_1)\dd F(t_2).\nonumber
\end{align}
Notice that the last term is equal to $A(v^*)$ defined in (\ref{p_fd_marginal}). 

Finally, consider the first two terms:
\begin{align*}
& (1-q)^{nm-m}+(n-1)\int_{t_1\geqslant t_2\geqslant v^*} F(t_2)^{nm-2m}\dd F(t_1)^{m-1}\dd F(t_2)^m \\
& = F(v^*)^{nm-m}+(m-1)\int_{v^*}^{\overline{v}} [F(t_1)^{nm-m}-F(v^*)^{nm-m}]F(t_1)^{m-2}\dd F(t_1)\\
& = F(v^*)^{nm-m}+ (m-1)\left[\frac{1-F(v^*)^{nm-1}}{nm-1}-\frac{F(v^*)^{nm-m}-F(v^*)^{nm-1}}{m-1}\right]\\
& = F(v^*)^{nm-m}+\frac{m-1}{nm-1}+\frac{m(n-1)}{nm-1}F(v^*)^{nm-1}-F(v^*)^{nm-m} = p(v^*),
\end{align*}
where $p(v^*)$ is defined in (\ref{p_nd_marginal}).

\subsection{Expected payoff for $v \geqslant v_{\rm F}^*$ increasing in $v$}

\label{sec:appendix_payoff_increasingFD}

Fix the marginal type to be $v_{\rm F}^*$ and let $q = 1-F(v_{\rm F}^*)$. For type $v \geqslant v_{\rm F}^*$, the expected payoff from entering is given by
\begin{align*}
& \Pi(v,v;v_{\rm F}^*) = v(1-q)^{nm-m} + (n-1)\sum_{k_1=0}^{m-1}\sum_{k_2=1}^m\sum_{k_3=0}^{nm-2m}\binom{m-1}{k_1}\binom{m}{k_2}\binom{nm-2m}{k_3}\times\\
&\times q^{k_1+k_2+k_3}(1-q)^{nm-1-k_1-k_2-k_3}(Y_1 + Y_2 + Y_3 + Y_4),
\end{align*} where
\begin{align*} 
& Y_1 = v \int\limits_{\overline{v} \geqslant t_1 > v} \int_{t_1\geqslant t_2\geqslant t_3 \geqslant v^*_{\rm F}}\left(1-\frac{t_2}{2t_1}\right)\dd\tilde{F}(t_1)^{k_1}\dd\tilde{F}(t_2)^{k_2}\dd\tilde{F}(t_3)^{k_3} \\
& Y_2 = v \int\limits_{\overline{v} \geqslant t_1 > v} \int_{t_2\geqslant t_1\geqslant t_3 \geqslant v^*_{\rm F}}\frac{t_1}{2t_2} \dd\tilde{F}(t_1)^{k_1}\dd\tilde{F}(t_2)^{k_2} \dd\tilde{F}(t_3)^{k_3} \\
& Y_3 = \int_{\overline{v} \geqslant v \geqslant t_1 \geqslant t_2 \geqslant t_3 \geqslant v^*_{\rm F}} (v - t_2)\dd\tilde{F}(t_1)^{k_1}\dd\tilde{F}(t_2)^{k_2} \dd\tilde{F}(t_3)^{k_3} \\
& Y_4 = \int_{\overline{v} \geqslant v \geqslant t_2 \geqslant t_1 \geqslant t_3 \geqslant v^*_{\rm F}} (v - t_1)\dd\tilde{F}(t_1)^{k_1}\dd\tilde{F}(t_2)^{k_2} \dd\tilde{F}(t_3)^{k_3}. 
\end{align*} 
This expression can be simplified considerably, first by integrating over $t_3$ and summing over $k_3$, then summing over $k_1$ and $k_2$ and swapping the variables of integration in two of the integral terms, which gives
\begin{align*}
& \Pi(v, v; v_{\rm F}^*) = v(1-q)^{nm-m} + m(m-1)(n-1) (vZ_1 + vZ_2 + vZ_3 - Z_4 - Z_5),
\end{align*}
where
\begin{align*}
& Z_1 = \int_{\overline{v}\geqslant t_1\geqslant t_2\geqslant v_{FD}^*} F(t_1)^{m-2} F(t_2)^{nm-m-1} \dd F(t_1) \dd F(t_2) \\
& Z_2 = \int\limits_{t_1 = v}^{\overline{v}} \int\limits_{t_2 = v_{FD}^*}^{t_1} \frac{t_2}{2t_1} F(t_1)^{m-2} F(t_2)^{nm-m-2} [F(t_1) - F(t_2)] \dd F(t_1) \dd F(t_2) \\
& Z_3 = \int_{v\geqslant t_1\geqslant t_2\geqslant v_{FD}^*} F(t_1)^{m-1} F(t_2)^{nm-m-2} \dd F(t_1) \dd F(t_2) \\
& Z_4 = \int\limits_{t_1 = v}^{\overline{v}} \int\limits_{t_2 = v_{FD}^*}^{t_1} t_1 F(t_1)^{m-1} F(t_2)^{nm-m-2} \dd F(t_1) \dd F(t_2) \\
& Z_5 = \int\limits_{t_1 = v}^{\overline{v}} \int\limits_{t_2 = v_{FD}^*}^{t_1} t_2 F(t_1)^{m-2} F(t_2)^{nm-m-1} \dd F(t_1) \dd F(t_2)
\end{align*}
It's straightforward to show that $vZ_1$, $vZ_3$ and $-Z_5$ are all increasing in $v$. Taking the derivative of $vZ_2 - Z_4$ with respect to $v$ gives 
\begin{align*}
& \frac{\dd(vZ_2 - Z_4)}{\dd v} = Z_2 - v\int\limits_{t_2 = v_{\rm F}^*}^{v} \frac{t_2}{2v} f(v) F(v)^{m-2} F(t_2)^{nm-m-2} [F(v) - F(t_2)] \dd F(t_2) + \\
& + f(v) \int\limits_{t_2 = v_{\rm F}^*}^{v} v F(v)^{m-1} F(t_2)^{nm-m-2} \dd F(t_2) \\
& = Z_2 - f(v) \int\limits_{t_2 = v_{\rm F}^*}^{v} \left[ \left(\frac{t_2}{2} - v \right) F(v)^{m-1} F(t_2)^{nm-m-2} - \frac{t_2}{2} F(v)^{m-2} F(t_2)^{nm-m-1} \right] \dd F(t_2).
\end{align*} 
Since $v \geqslant t_2$, the integrand is everywhere negative, which ensures that $vZ_2 - Z_4$ is increasing in $v$. It follows that $\Pi(v, v; v_{\rm F}^*)$ is increasing in $v$ for $v \geqslant v_{\rm F}^*$, as desired.

\subsection{Equation (\protect\ref{B_fd}) and proof of Proposition \protect\ref{prop_comp_nd_fd}}

\label{sec:appendix_Eq15}

Start by summing up over $k_1$, $k_2$ and $k_3$ in (\ref{B_fd_def}):
\begin{align*}
& B_{\rm F} = n(n-1)\int_{t_1\geqslant t_2\geqslant t_3\geqslant v^*_{\rm F}}\left(\frac{t_2}{2}+\frac{t_2^2}{2t_1}\right)F(t_2)^{nm-2m}\dd F(t_1)^m\dd F(t_2)^m \\
& = m^2n(n-1)\int_{t_1\geqslant t_2\geqslant t_3\geqslant v^*_{\rm F}}\left(\frac{t_2}{2}+\frac{t_2^2}{2t_1}\right)F(t_1)^{m-1}F(t_2)^{nm-m-1}\dd F(t_1)\dd F(t_2).
\end{align*}
Subtracting aggregate investment in the case of no disclosure, Eq. (\ref{B_nd}), obtain
\begin{align*}
& \frac{B_{\rm F}-B_{\rm N}}{m^2n(n-1)} = \int_{t_1\geqslant t_2\geqslant v^*_{\rm F}}\left(\frac{t_2}{2}+\frac{t_2^2}{2t_1}\right)F(t_1)^{m-1}F(t_2)^{nm-m-1}\dd F(t_1)\dd F(t_2)\\
& - \int_{t_1\geqslant t_2\geqslant v^*}t_2F(t_1)^{m-1}F(t_2)^{nm-2}\dd F(t_1)\dd F(t_2)\\
& \geqslant \int_{t_1\geqslant t_2\geqslant v^*}t_2F(t_2)^{nm-m-1}\left[\left(\frac{1}{2}+\frac{t_2}{2t_1}\right)F(t_1)^{m-1}-F(t_2)^{m-1}\right]\dd F(t_1)\dd F(t_2)\\
&\geqslant \int_{t_1\geqslant t_2\geqslant v^*}t_2F(t_2)^{nm-m-1}\left[\frac{t_2}{t_1}F(t_1)^{m-1}-F(t_2)^{m-1}\right]\dd F(t_1)\dd F(t_2).
\end{align*}
The first inequality follows from the fact that $v^*_{\rm F}\leqslant v^*$ and $F(t_1)\leqslant 1$, and the second is obtained by replacing $\frac{1}{2}$ with $\frac{t_2}{2t_1}\leqslant \frac{1}{2}$.

Note that $t_1\geqslant t_2$ everywhere in the domain of integration. Thus, in order to show that the expression above is positive it suffices to show that $\frac{F(t)^{m-1}}{t}$ is increasing in $t$, which is equivalent to the assumption on the elasticity of $F(\cdot)$ in the proposition.

\subsection{Two groups with different sizes for $F(v)=v^{\alpha}$}
\label{sec:appendix_hetero}

Consider the distribution of values $F(v)=v^{\alpha}$ on $[0,1]$ and assume $m_1\leqslant m_2$. The results in Section \ref{sec:hetero_sizes} then give the following.

Under ND,
\begin{align*}
& \gamma_{\rm N}(v) = \left(\frac{m_1}{m_2}v^{\alpha-1}+1-\frac{m_1}{m_2}\right)^{1/(\alpha-1)},\\
& B_{1,{\rm N}} = \int_0^1 \gamma_{\rm N}(s)^{1+\alpha(m_2-1)}(1-s^{\alpha m_1})\dd s^{\alpha m_1},\\
& B_{2,{\rm N}} = \int_{\gamma_{\rm N}(0)}^1 \gamma_{\rm N}^{-1}(s)^{1+\alpha(m_1-1)}(1-s^{\alpha m_2})\dd s^{\alpha m_2}.
\end{align*}
Note that (i) For $\alpha=1$, we can continuously define $\gamma_{\rm N}(v)=v^{m_1/m_2}$; (ii) $\gamma_{\rm N}^{-1}$ can be found by swapping $m_1$ and $m_2$ in $\gamma_{\rm N}$; and (iii) $\gamma_{\rm N}(0)=0$ for $\alpha\le 1$, whereas $\gamma_{\rm N}(0)>0$ for $\alpha>1$.

Under WD,
\begin{align*}
& \gamma_{\rm W}(v) = \left[1+\frac{m_1(\alpha m_2-1)}{m_2(\alpha m_1-1)}(v^{\alpha m_1-1}-1)\right]^{1/(\alpha m_2-1)},\\
& B_{1,{\rm W}} = \int_{\gamma_{\rm W}^{-1}(0)}^1\gamma_{\rm W}(s)(1-s^{\alpha m_1})\dd s^{\alpha m_1}, \\
& B_{2,{\rm W}} = \int_{0}^1\gamma_{\rm W}^{-1}(s)(1-s^{\alpha m_2})\dd s^{\alpha m_2}.
\end{align*}
The expression for $\gamma_{\rm W}$ is valid assuming $\alpha m_1,\alpha m_2\ne 1$, and we omit the corresponding limiting cases for brevity. As above, $\gamma_{\rm W}^{-1}$ can be found by swapping $m_1$ and $m_2$ in $\gamma_{\rm W}$.

Finally, under FD we obtain
\begin{align*}
B_{\rm F} = \frac{\alpha^2 m_1m_2}{2(\alpha m_1+\alpha m_2+1)}\left[\frac{2\alpha m_1+3}{(\alpha m_1+1)(\alpha m_1+2)} + \frac{2\alpha m_2+3}{(\alpha m_2+1)(\alpha m_2+2)}\right].
\end{align*}

\subsection{Proof of Proposition \protect\ref{prop_comp_nd_fd_max}}

\label{sec:appendix_proofProp5}

By integrating over $t_1$, Eq. (\ref{B_nd_max}) can be simplified as follows:
\begin{align*}
& B^{\max}_{\rm N} = m(n-1)\int t_2F(t_2)^{nm-2}[1-F(t_2)^{nm}]\dd F(t_2) \\
& = \frac{m(n-1)}{nm-1}\int t\dd F(t)^{nm-1} - \frac{m(n-1)}{2nm-1}\int t\dd F(t)^{2nm-1}.
\end{align*}
When either $n\to\infty$ or $m\to\infty$ (or both), each of the integrals converges to $\overline{v}$. The coefficients in front of the integrals converge to 1 and $\frac{1}{2}$, respectively, when $n\to\infty$ for a fixed $m$; and to $\frac{n-1}{n}$  and $\frac{n-1}{2n}$, respectively, when $m\to\infty$ for a fixed $n$. Thus, $B^{\max}_{\rm N}$ converges to $\frac{\overline{v}}{2}$ when $n\to\infty$ for a fixed $m$; and to $\frac{(n-1)\overline{v}}{2n}$ when $m\to\infty$ for a fixed $n$. 

A lower bound for $B^{\max}_{\rm F}$ can be obtained by replacing $t_1$ with $\overline{v}$ in Eq. (\ref{B_fd_max}). Integrating over $t_1$ similar to the above, we obtain
\begin{align*}
& B^{\max}_{\rm F} > mn(n-1)\int \left(\frac{t_2}{2}+\frac{t_2^2}{6\overline{v}}\right)F(t_2)^{nm-m-1}[1-F(t_2)^m]\dd F(t_2) \\
& = n\int \left(\frac{t}{2}+\frac{t^2}{6\overline{v}}\right)\dd F(t)^{(n-1)m} - (n-1)\int \left(\frac{t}{2}+\frac{t^2}{6\overline{v}}\right)\dd F(t)^{nm}.
\end{align*}
When $m\to\infty$, for $n$ fixed, each of the integrals converges to $\frac{2\overline{v}}{3}$, and hence the lower bound of $B^{\max}_{\rm F}$ converges to $\frac{2\overline{v}}{3}$ as well.

Consider now the limit $n\to\infty$, for $m$ fixed. For brevity, let $\gamma(t)=\frac{t}{2}+\frac{t^2}{6\overline{v}}$. We can then write
\begin{align}
\label{bound_os}
B^{\max}_{\rm F} > n\mathds{E}(\gamma(Y_{(n-1:n-1)}))-(n-1)\mathds{E}(\gamma(Y_{(n:n)})),
\end{align}
where $Y_{(r:n)}$ are the order statistics from a sample of i.i.d. random variables distributed according to $F(\cdot)^m$. Note that $\gamma(\cdot)$ is a polynomial function; therefore, we can apply the recurrence relation for moments of order statistics \citep[][Section 3.4]{David-Nagaraja:2003}:
\[
(n-r)\mathds{E}(\gamma(Y_{(r:n)})) + r\mathds{E}(\gamma(Y_{(r+1:n)})) = n\mathds{E}(\gamma(Y_{(r:n-1)})).
\]
For $r=n-1$, this gives 
\[
\mathds{E}(\gamma(Y_{(n-1:n)})) + (n-1)\mathds{E}(\gamma(Y_{(n:n)})) = n\mathds{E}(\gamma(Y_{(n-1:n-1)})), 
\]
and (\ref{bound_os}) then implies
\[
B^{\max}_{\rm F} > \mathds{E}(\gamma(Y_{(n-1:n)})),
\]
where the right-hand side converges to $\frac{2\overline{v}}{3}$. This establishes the claim.

\end{document}